\documentclass{kapedbk} 
\usepackage[dvips]{graphicx}
\chaptersection 
\upperandlowercase
\let\footnote\savefootnote

\kluwerbib 

\begin{document}
\articletitle{A panchromatic view of AGN}

\rhead{A panchromatic view of AGN}

\author{Guido Risaliti \& Martin Elvis}

\affil{Harvard-Smithsonian Center for Astrophysics\\
60 Garden Street, Cambridge, MA 02138}
\email{grisaliti,melvis@cfa.harvard.edu}

\begin{abstract}
We review the continuum emission of Active Galactic Nuclei (AGN) 
over the entire electromagnetic spectrum. After a brief historical 
introduction, we describe the main spectral properties of unobscured 
AGN, discussing the selection biases which prevent us from having 
a complete view of the AGN population in the universe, and trying 
to build an updated spectral energy distribution of optically 
selected quasars. In the second part of the review, we describe 
the spectral properties of obscured AGN. Finally, we discuss the 
main observational methods in the different wavelength bands
for disentangling AGN and stellar emission, and the ability of
these methods to find new (mainly obscured) AGN, a significant 
fraction of which are probably still missing in current surveys.
\end{abstract}

\section{Introduction}

Active Galactic Nuclei (AGN) shine over $\sim20$ decades of the 
electromagnetic spectrum, from the radio to the gamma rays. 
In almost all of this huge energy range, AGN are the brightest 
sources in the sky, except for the relatively narrow 
($\sim$3 decades) range from the infrared to the ultraviolet (UV).

The ``modern history'' of quasars is closely linked with the 
development of non-optical astronomy. In the early 60s, hard won 
precise positions of radio sources enabled the large optical 
telescopes of the day to take spectra of whatever optical object 
lay at that position. Some showed the starlight of normal galaxies, 
but often at large distances. In 1963 Marteen Schmidt made the 
``official'' discovery of the first quasar, 3C~273, realizing that 
the point-like counterpart of a powerful radio source was at a 
redshift $z=0.16$, implying an enormous luminosity.
In the early sixties, the availability of new space-borne instruments 
also opened the new field of astronomical observations in the X-rays, 
which soon revealed a sky dominated by AGN.  

It is now clear that the initial population of quasars discovered 
in radio surveys is only a small fraction of a 10 times more numerous 
class of quasars, most of which are ``radio quiet''. These sources 
are, in turn, a small fraction of the total AGN population, which 
is dominated by obscured sources that can only be detected through 
their hard X-ray emission or through their reprocessed radiation 
in the infrared.

In this chapter, we describe the main continuum properties of 
AGN across the whole electromagnetic spectrum,
according to our present, still incomplete, knowledge. 
We do not treat the emission of the small minority of quasars known 
as ``blazars'', whose emission appears significantly altered by 
relativistic beaming effects. We focus on the 
{\em observed\/} properties of quasars from a phenomenological 
point of view. More emphasis on the physical processes responsible 
for the observed emission of quasars can be found in
Armitage (this volume).

We start in \S\ref{risalitisectype1} with unobscured
(type I) AGN. We briefly discuss the main limitations in 
defining a representative sample of AGN. We then focus 
on a sample of bright, optically selected quasars (the Bright 
Quasar Survey, Schmidt \& Green 1983) in order to build a mean 
Spectral Energy Distribution (SED) that takes into account all 
of the most recent observational results. This can be 
considered an update of the 10 year old compilation of 
Elvis et al.\ (1994; hereafter, E94).
We conclude \S\ref{risalitisectype1} with a brief overview of 
the possible dependence of the average spectral properties of 
type I AGN on redshift and/or luminosity and with a brief 
discussion on the often underestimated issue of the intrinsic
dispersion of quasar spectra with respect to their average shapes.

Section~\ref{risalitisecseds} is devoted to obscured (type II) 
AGN. We first review the main spectral 
properties for each energy band. We then discuss the relation 
between dust and gas absorption, describing some recently 
discovered sources for which the ratio between gas 
and dust, and the dust properties, are strongly different from 
those in our Galaxy. Finally, in \S\ref{risalitisectype2} we 
discuss the current methods that can be used to discover obscured 
AGN in the different energy bands. There are several indications 
that a significant fraction of these sources are still missing 
in current surveys. In particular, we focus on the 
techniques that can help to disentangle the AGN and stellar 
emission in the population of luminous sources that dominate 
the infrared and submillimeter sky. We conclude with a brief 
summary in \S\ref{risalitisecsummary}.

We finish this introduction with a note on terminology:
historically, the term ``quasar'' has been used widely to 
refer to AGN of high luminosity whose emission completely 
overwhelmed that of the host galaxy. (Thus, these sources 
appeared as point sources in optical observations.) Low-luminosity 
AGN were (and still are) referred to as ``Seyfert Galaxies''.
However, evidence for any basic physical difference between 
these types of active objects has diminished through the 
years, essentially to the vanishing point. For this reason, 
in this chapter, we will use the term ``quasar'' as synonymous 
with type 1 AGN. 

\section{Spectral Properties of Quasars}
\label{risalitisectype1}

Defining the SED of quasars is an extremely difficult task 
from an observational point of view. This is mainly due to 
observational biases introduced by the narrow wavelength ranges 
used for the selection of all quasar samples.

Historically, the most common technique for quasar selection 
has been the UV-blue excess in the quasar continuum. For example, 
the best-studied quasar sample, the Bright Quasar Survey sample
(part of the Palomar-Green survey, Schmidt \& Green 1983, and 
hereafter referred to as ``PG quasars''), is defined through a 
magnitude limit ($B<16.2$, $M_V<-23$) and a color limit 
($U-B<-0.4$). Detailed studies of this sample of quasars
revealed a remarkable
homogeneity in their continua (Sanders et al.\ 1989; E94) 
and line properties from the infrared to the X-rays (Laor et al.\ 1997), 
with the exception of a small fraction ($\sim$10\%) of Broad 
Absorption Line (BAL) objects with quite different continua and line
properties. Subsequent quasar samples have been obtained with deeper 
surveys but similar selection criteria.

%
%
\begin{figure}[hbt]
\centerline{\includegraphics[width=10cm]{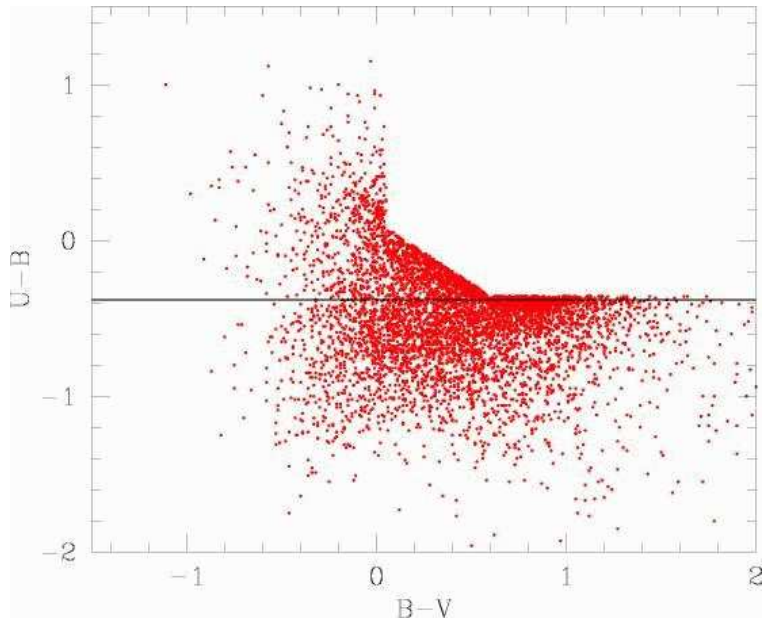}}
\caption{Color-color diagram for 2dF quasars. Horizontal line shows 
the ``classical'' selection criterion for blue quasars ($U-B<-0.4$).
Empty region in the upper right is dominated by stars, and has 
therefore been excluded in the 2dF quasar search.}
\label{risaliti_colors}
\end{figure}

However, the blue selection method is in no way complete and can 
provide only a partial view of the general quasar SED. This is 
evident in Figure~\ref{risaliti_colors}, which shows the location 
of quasars from the Anglo-Australian Telescope's Two Degree Field 
(2dF) QSO Redshift Survey (Croom et al.\ 2002) in the ($B-V$, $U-B$)
plane. The ``classical'' UV excess selection criterion 
{\em (horizontal line)\/} was chosen for efficiency 
(selecting objects that are quasars) and not for completeness 
(selecting all quasars). 
Hence, the selection limit is in no way related to any intrinsic 
property of the quasars, but rather to the UV emission of stars: 
the color limit is chosen in order to reject the great majority of 
stars and thus have a high efficiency in quasar selection. 
In principle, a more relaxed color selection would find more quasars, 
but the fraction of stars among the selected objects would be much 
higher, making the spectroscopic follow-up extremely time-consuming.

Many other criteria have been adopted in order to increase the 
completeness. At other wavelengths, radio surveys like the Very Large 
Array (VLA)'s Faint Images of the Radio Sky at Twenty-cm (FIRST)
Bright Quasar Survey (White et al.\ 2000) are now deep enough
to find ``radio-quiet'' objects (which are much fainter than radio-loud
quasars at radio wavelengths, but are still detectable). X-ray 
surveys are also effective at discovering AGN that would be missed 
in optical searches because of their red colors or because of 
absorption (Brusa et al.\ 2003; Barger et al.\ 2003). 
In the optical band, spectroscopic criteria have been 
used to discover quasars from their broad emission lines
instead of their continuum properties, 
e.g., the Palomar-Grism survey (Schneider et al.\ 1994),
the Hamburg Quasar Survey (Hagen et al.\ 1995), and the 
Hamburg-ESO survey (Wisotzki et al.\ 1997). Finally, multicolor 
optical selections have been used in recent surveys, like the 
Sloan Digital Sky Survey (SDSS). 

In the following, we will focus on 
blue-selected quasars in order to build a quasar SED. The reason 
for this choice is twofold: (i) blue-selected quasars are currently 
the best-studied, i.e., the ones with the best available
multiwavelength observational data, and (2) their emission is thought 
to be representative of the {\em intrinsic\/} emission of most quasars. 
The cases for intrinsically different SEDs and for absorbed objects 
will be discussed later.

The standard reference for quasar SEDs is the atlas of E94,
who collected observations of bright quasars from radio to hard 
X-ray wavelengths. The mean SED from this work is often cited as 
the ``standard'' quasar SED, and the main properties of this SED 
are in good agreement with the SEDs of optically selected quasars, 
such as the PG quasars. However, even within this sample, the
90\% range of SEDs spans factors of $\sim10$ or more at most 
frequencies. There are three main limitations to the E94 atlas:\\

\noindent
1) {\em Selection criterion.\/} The E94 sample is defined by requiring 
the presence of a detection sufficient to yield a spectrum in an 
{\em Einstein Observatory\/} X-ray observation. As discussed in E94, 
this introduces a well known bias towards X-ray bright objects. For 
comparison, the average optical-to-X-ray flux ratio
index\footnote{$\alpha_{OX}$ is defined as the slope
of a nominal power law connecting the continuum at 2500~\AA\ with
that at 2~keV
($\alpha_{OX}=0.385\times\log(f(2500{\rm\AA})/f(2~{\rm keV}))$},
$\alpha_{OX}$, is 1.35 for the E94 sample and 1.55
for the low redshift PG quasars (Laor et al.\ 1997), corresponding 
to a factor of 3 higher X-ray emission in the E94 sample.  \\

\noindent
2) {\em Lack of data.\/} The E94 sample did not have data in the range 
between the Lyman edge (13.2~eV) and the soft X-rays ($0.2-0.4$~keV).\\

\noindent
3) {\em Limited number of quasars.\/} The E94 sample had 29 
radio-quiet quasars and 18 radio-loud.\\

The first two points can now be addressed well, and a significant 
improvement is also possible for the third point. It is possible to 
quantitatively discuss the effects of the first point using 
observations at different wavelengths to estimate the correct ratios
between the emissions in the different bands. Regarding the second 
point, a big part of the gap between the Lyman edge and the soft 
X-rays can now be filled, thanks to the availability of 
{\em Hubble Space Telescope (HST)\/} and 
{\em Far Ultraviolet Spectroscopic Explorer (FUSE)\/} UV spectra 
of high redshift quasars.

In the following, we build an average SED using mainly---but 
not only---the best studied quasars, i.e., the local ($z<0.4$) PG 
quasars ($B<16.2$, $U-B<-0.4$). We adopt a slightly different 
approach with respect to E94: in order to produce a final average
SED, we work out the average spectrum in each
observational band, and then we estimate the bolometric 
corrections for each spectral region.
We make use of the latest observational results, in particular,
those from the {\em Infrared Space Observatory (ISO)\/} in 
the infrared (Haas et al.\ 2001), {\em HST\/} in the UV for quasars 
(Telfer et al.\ 2002) and bright, lower luminosity Seyfert 1
galaxies (Crenshaw et al.\ 1999), and {\em ROSAT\/}, {\em BeppoSAX\/}, 
and {\em ASCA\/} in the X-rays (Laor et al.\ 1997; Mineo et al.\ 2000; 
George et al.\ 2000).

\subsection{Optical/UV} 

The optical to UV emission of quasars is characterized by the 
``big blue bump'' (Shields 1978; Malkan \& Sargent 1982; Elvis 1985),
where the peak of quasar emission is usually found. The peak energy 
is around the Lyman edge ($\lambda=1216$~\AA), and the spectrum can 
be well approximated with a power law both at lower and higher 
frequencies. Recent observations performed with {\em HST\/} of over 
200 quasars (compiled by Telfer et al.\ 2002) provide a good
quality mean spectrum from $\lambda\sim300$~\AA\ to $\lambda\sim3000$~\AA.
Composite spectra (see Fig.~\ref{risaliti_composite})
extending from $\lambda\sim1200$~\AA\ to 
$\lambda\sim9000$~\AA\ have been obtained using data from
ground-based optical surveys like the UK Schmidt Telescope's 
Large Bright QSO Survey (Francis et al.\ 1991),
the 2dF (Croom et al. 2002), the optical follow-up of the 
FIRST radio survey (Brotherton et al.\ 2001), and the 
SDSS (Vanden Berk et al.\ 2001). The SDSS includes more than
2200 spectra at redshifts between 0 and $\sim 5$, providing the most
accurate average optical spectrum of quasars so far, with a spectral 
resolution of a few \AA. In the overlapping band ($\sim1200-3800$~\AA), 
the {\em HST\/} and SDSS results match within the errors,
providing a complete quasar spectrum in the $300-9000$~\AA\ band.
The main results of these studies are the following:

%
%
\begin{figure}[hbt]
\centerline{\includegraphics[width=11.5cm]{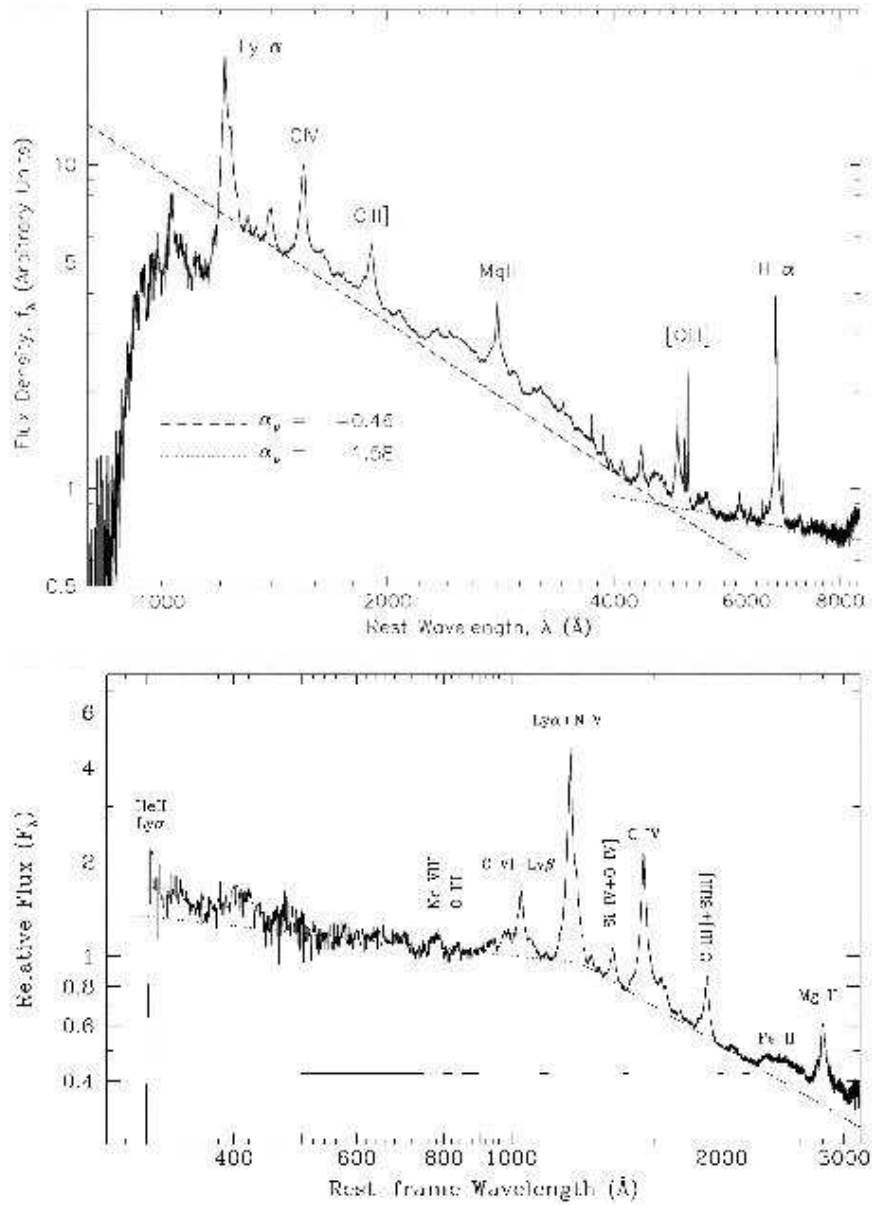}}
\caption{Composite optical/UV spectra of quasars.
{\em (Upper panel)\/} Results from the SDSS (Vanden Berk et al.\ 2001;
their Fig.~3).
{\em (Lower panel)\/} Results from {\em HST\/} (Telfer et al.\ 2002;
their Fig.~4). Dotted and dashed lines are power law fits to the continuum.
Horizontal thick lines in the lower panel show the spectral regions
used to estimate the continuum level.
}
\label{risaliti_composite}
\end{figure}

$\bullet$ The $300-5000$~\AA\ continuum can be modeled with two power 
laws with slopes\footnote{Somewhat different conventions are used for 
the power law index: photon index $\Gamma$ in the X-rays 
(counts/sec/keV $\propto E^{-\Gamma}$), $\alpha$ in the radio 
(f$_\nu \propto \nu^{\alpha}$), and $-\alpha$ or $\alpha_\lambda$ 
in the optical. In this review, we follow the radio convention, as 
it is mathematically correct in the $\log f_\nu$ versus $\nu$ space.}
$\alpha_1=-1.76$ 
between 300~\AA\ and $\sim1200$~\AA\ and $\alpha_2=-0.44$ between 
1200~\AA\ and 5000~\AA. At longer wavelengths, the spectrum appears 
to flatten significantly, but this effect is probably due to the 
contribution of galactic emission in the quasar spectra 
(see Vanden Berk et al.\ 2001 for more details).

$\bullet$ No correlation between optical continuum properties and 
redshift or luminosity has been found. An anticorrelation between 
the equivalent width (EW) of the main emission lines and the luminosity 
(the ``Baldwin effect'', Baldwin 1977) has been found.

$\bullet$ Hundreds of emission lines are present in quasar spectra 
(a compilation of the brightest ones is given in 
Table~\ref{risaliti_lines}). In addition to these lines,
another major feature in the optical/UV spectra of quasars is the 
``small blue bump'' (Wills, Netzer, \& Wills 1985; Elvis 1985) between 
$\sim2200$~\AA\ and $\sim4000$~\AA. This is not a true continuum 
feature but is due to a forest of emission lines from the ion FeII
and the Balmer recombination continuum. Permitted emission lines are 
``broad'' (corresponding to velocities of the emitting gas of
$2000-15000$~km~s$^{-1}$), while forbidden lines are narrow (a few 
hundred km~s$^{-1}$). 

$\bullet$ A minority ($\sim10-20$\%) of quasars show broad absorption 
lines, often saturated, with widths and blueshifts of several 
thousand km~s$^{-1}$ and peaks of $20,000-30,000$~km~s$^{-1}$. 

$\bullet$ $\sim50$\% of bright Seyfert 1 galaxies observed in the UV
with {\em HST\/} or {\em FUSE\/} show evidence of Narrow Absorption Lines 
(NAL), with widths of $\sim1000$~km~s$^{-1}$, in the profiles of
high ionization emission lines (OVI, CIV, Ly$\alpha$).
The presence of such features is strongly correlated 
with the presence of warm absorbers in the soft X-rays 
(see \S\ref{risalitisecxrays}).

\begin{table}[tbh]
\caption[Main emission lines in the optical/UV
spectrum of quasars]
{Compilation of the emission lines with EW$>$5~\AA\ in the
optical/UV spectra of quasars}
\centering
\label{risaliti_lines}
\begin{tabular}{lccl|lccc}
\sphline
\em Line & $\lambda$(\AA) & $\langle {\rm EW(\AA)}\rangle$& \em Ref. &\em Line
 & $\lambda$(\AA) & $\langle {\rm EW(\AA)}\rangle$
& \em Ref.\\
\sphline
CIII+NIII     & 980  & 9.7$\pm0.2$  & (1) &MgII          & 2799 & 32.3$\pm$0.1
& (2) \\
OVI+Ly$\beta$ & 1030 & 15.6$\pm0.3$ & (1) & H$\delta$     & 4103 & 5.05$\pm$0.06 & (2) \\
Ly$\alpha$    & 1216 & 91.8$\pm0.7$ & (1) & H$\gamma$     & 4341 & 12.6$\pm$0.1& (2) \\
N~V            & 1240 & 18.5$\pm0.5$ & (1) & H$\beta$      & 4863 & 46.2$\pm$0.2& (2) \\
SiIV+OIV]     & 1397 & 8.13$\pm$0.09& (2) & $[$OIII]       & 5007 & 13.2$\pm$0.2& (2) \\
CIV           & 1549 & 23.8$\pm$0.1 & (2)$^a$ &H$\alpha$     & 6565 & 194.5$\pm$0.& (2) \\ CIII]         & 1909 & 21.2$\pm$0.1 & (2) \\
\sphline
\end{tabular}
\begin{tablenotes}
References --- (1) Telfer et al.\ 2002 (for lines at
$\lambda<1300$~\AA); (2) Vanden Berk et al.\ 2001
(for $\lambda>1300$~\AA).
\\
\smallskip
Table notes --- $^a$This is the only line in the $1300-3000$~\AA\
range for which the two references above give discrepant values
(Vanden Berk reports the EW(CIV$\lambda1549)=23.8\pm 0.1$ reported).
\end{tablenotes}
\end{table}

\subsection{Radio/submillimeter}

The radio emission of PG quasars (Kellerman et al.\ 1989) is 
significantly different for radio-loud and radio-quiet sources.
However, in all cases, the radio emission provides a negligible 
fraction of the 
bolometric luminosity. Here we only concentrate on the ``core'' 
emission, i.e., the flat spectrum, compact component that is 
physically distinct from the steep spectrum lobes (however, 
the angular resolution is not always good enough to separate 
these components).  In radio-loud objects, a strong, non-thermal
continuum extends from the radio to the far-infrared 
through the submillimeter, while in radio-quiet objects, the 
SED turns over sharply in the far-infrared, with a slope $\alpha>2.5$ 
indicative of dust, and the radio emission is only a negligible 
tail of this component.

\subsection{Infrared} 

The infrared (IR) emission of PG quasars has been systematically 
studied with the {\em Infrared Astronomical Satellite\/} 
({\em IRAS\/}; Sanders et al.\ 1989), and, more recently, with the 
{\em ISO\/} satellite (Haas et al.\ 2003). The latter work confirms 
the basic results of the former, while adding further details. 
The basic characteristics of the IR emission of quasars are the 
following:

$\bullet$ The integrated IR emission ($2-200\mu$m) is, on average, 
$\sim30$\% of the bolometric luminosity, with values in individual 
objects ranging from $\sim15$\% to $\sim50$\%.
The spectral shape is characterized by (i) a minimum at $\sim1-2\mu$m, 
corresponding to the sublimation temperature of the most refractary 
dust (between 1000 and 2000~K, depending on the composition
of the dust grains), (ii) an ``IR bump'', typically at $10-30\mu$m 
(but there are examples of flat spectra, or peaks anywhere between 
2 and $100\mu$m), due to the thermal emission of dust, with a 
temperature range between 50 and 1000~K, and (iii) a steep decline 
($f_\nu \propto \nu^{\alpha}, \alpha>3$) at large wavelengths, 
typical of the low energy spectrum of a gray emitter 
(Chini et al.\ 1989).

$\bullet$ The spectral shape of most of the sources in the sample 
is better reproduced, according to Haas et al.\ (2003), by 
reprocessing of the quasar primary emission, with the contribution 
of a starburst being negligible. However, this is still a 
controversial point, since the IR continuum expected from a quasar 
or a starburst is strongly dependent on the geometric and physical 
properties of the reprocessing medium, and the same observed 
continuum can often be successfully explained with more than one 
model (Elitzur, Nenkova, \& Ivezic 2004).

$\bullet$ The far-IR emission of radio-loud quasars is quite different 
than that of radio-quiet quasars. The spectrum between the IR bump 
and the submillimeter range is well reproduced by a power law with 
spectral index close to $\alpha=2.5$, as expected from self-synchrotron
absorption. The main emission mechanism here is not reprocessing by 
dust, but synchrotron emission by relativistic electrons.

\subsection{X-rays} 
\label{risalitisecxrays}

\begin{figure}[hbt]
\centerline{\includegraphics[width=12cm]{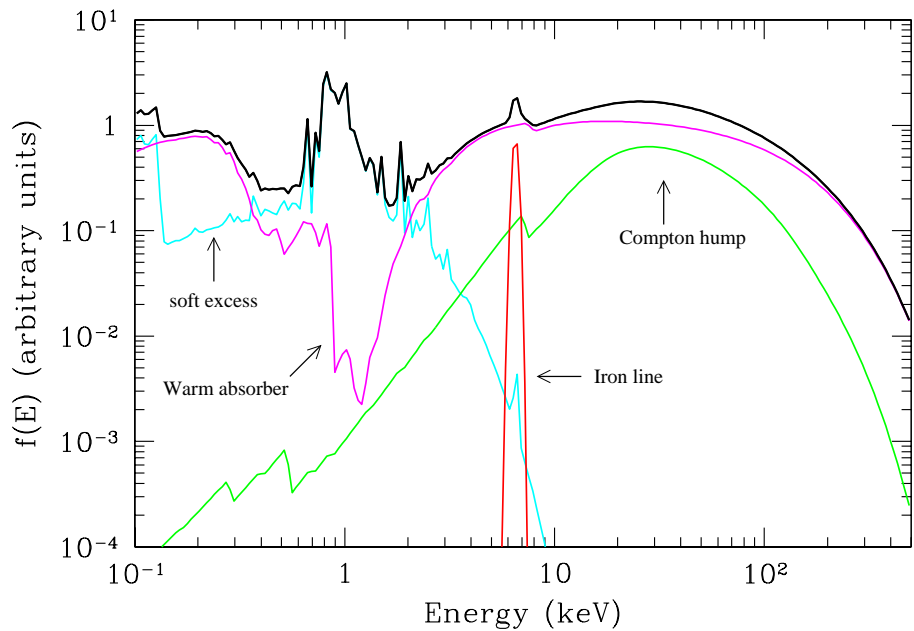}}
\caption{Average total spectrum {\em (thick black line)\/} and main 
components {\em (thin grey lines)\/} in the X-ray spectrum of a type I AGN.
The main primary continuum component is a power law with an high energy
cut-off at E$\sim100-300$~keV, absorbed at soft energies by warm
gas with $N_H\sim10^{21}-10^{23}$~cm$^{-2}$. A cold reflection 
component is also shown. The most relevant narrow feature is the iron
K$\alpha$ emission line at 6.4~keV. Finally, a ``soft excess'' is shown, 
due to thermal emission of a Compton thin plasma with temperature
$kT\sim0.1-1$~keV.
}
\label{risaliti_xraymodel}
\end{figure}

The X-ray properties of bright, optically selected quasars have 
been intensively studied in the last 25 years (Elvis et al.\ 1978; 
Zamorani et al.\ 1981), mostly with broadband but low resolution 
spectra. The X-ray emission from quasars extends from the Galactic 
absorption cut-off at $\sim0.1$~keV up to $\sim300$~keV. 
Laor et al.\ (1997) analyzed the {\em ROSAT\/} soft X-ray ($0.5-2$~keV)
observations of the sample of local ($z<0.4$) PG quasars, and a 
subsample of these objects has been studied with {\em ASCA\/} in the 
$2-10$~keV band (George et al.\ 2000) and with {\em BeppoSAX\/} in
the $1-100$~keV band (Mineo et al.\ 2001). Recent studies of samples 
of bright Seyfert 1 galaxies are reported in George et al.\ (1997;
{\em ASCA\/} observations) and in Perola et al.\ (2002; BeppoSAX 
observations). The main properties of the X-ray spectra of 
type I AGN are briefly summarized below and are shown in 
Figure~\ref{risaliti_xraymodel}.

$\bullet$ {\em Primary emission.\/}
The intrinsic continuum X-ray emission of quasars is to first 
order a power law, extending from about 1~keV to over 100~keV. 
However, as higher resolution and better signal-to-noise spectra 
have become available, emission and absorption features have been 
found that mask a direct view of this ``power law'' over virtually 
the whole X-ray band (see Fig.~\ref{risaliti_xraymodel}).
Hence, slight curvatures may be present but unseen. The typical 
spectral index\footnote{X-ray astronomers tend to use
the ``photon index'' $\Gamma$, where $N(E)\propto E^{-\Gamma}$ and
$\alpha=-(\Gamma-1)$.} is between $\alpha=-0.8$ and
$\alpha=-1$, both for low luminosity Seyfert galaxies 
and high luminosity quasars.
Radio-loud AGN have a somewhat flatter spectrum ($\alpha$ 
between $-0.5$ and $-0.7$). This is thought to be due to the 
additional hard component emitted by inverse-Compton scattering
of the electron in the jet on the radio-synchrotron photons, 
but this is not fully established.

There is now increasing evidence, mostly from {\em BeppoSAX\/}, 
for a roughly exponential cut-off to the power law at energies 
$\sim 80-300$~keV. This is presumably due to the cut-off in the 
energy distribution of the electrons responsible for the X-ray emission.
It is still debated whether the spectral index is redshift or 
luminosity dependent (Zamorani et al.\ 1981; Avni \& Tananbaum 1982, 1986; 
Bechtold et al.\ 2002; Vignali et al. 2003).

%
%

In addition to the main power law continuum component, a soft 
emission component is often observed in AGN, with characteristic 
temperature kT$\sim 0.2-1$~keV. The physical origin of this component
is not clear: warm emitting gas could be located in the accretion
disk, or in the broad line region (it could be the confining medium 
of the broad emission line clouds), or in a region farther from the 
center. Alternatively, this ``soft excess'' could be an extension 
of the big blue bump to higher energies, e.g., via Compton scattering 
in a hot accretion disk corona (Czerny \& Elvis 1987).

$\bullet$ {\em Reflection components.\/}
The primary emission of AGN can be ``reflected'', i.e., Thomson 
scattered by ionized gas. If the reflector has a column density 
$N_H>1.5\times10^{24}$~cm$^{-2}$ (i.e., $\sim1/\sigma_{T}$)
and is not fully ionized, the reflected component has a spectrum 
like the one shown in Figure~\ref{risaliti_xraymodel} (the actual 
shape slightly varies, depending on the geometry and chemical 
composition of the reflector). The main features of this reflection 
component are a continuum due to electron scattering
with a peak at $\sim30$~keV, and a cut-off at $4-5$~keV due to 
photoelectric absorption of the lower energy incident radiation. 
The reflection efficiency is typically a few percent of the direct 
emission in the $2-10$~keV range because of photoelectric
absorption, rising to $\sim30$\% at the 30~keV peak for a Compton-thick
reflector covering a significant fraction of the solid angle 
(Ghisellini et al.\ 1994). The efficiency drops if the reflecting 
medium is Compton thin (in this case part of the incident radiation 
escapes without interaction).

A warm, ionized reflector must be present in the central region of 
many AGN (since we see a ``warm absorber'' in $\sim$50\% of Seyfert 1
galaxies). The reflected emission has the same spectral shape as the 
incident continuum.

%
%
\begin{figure}
\centerline{\includegraphics[width=11.5cm]{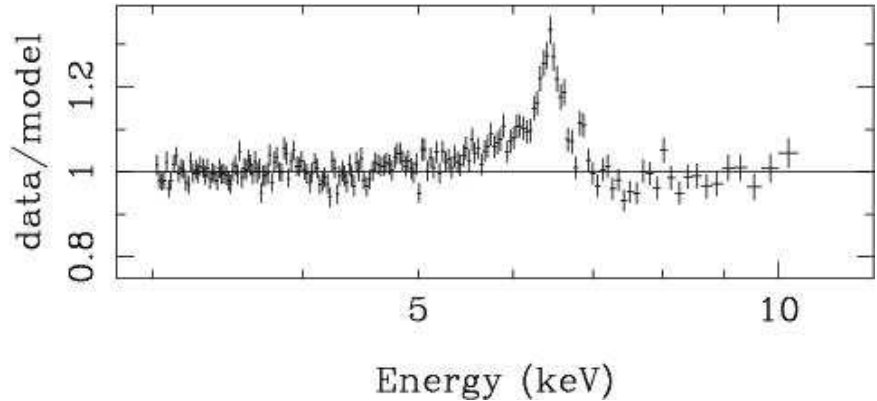}}
\caption{Iron line profile in an {\em XMM-Newton\/} observation of the
Seyfert 1 galaxy MCG-6-30-15. Model is a power law fitted in the
$3-5$~keV and $8-10$~keV energy bands.
Figure is from Vaughan \& Fabian (2004; their Fig.~8)}
\label{risaliti_ironline}
\end{figure}

$\bullet$ {\em Iron line.\/}
The most prominent narrow feature in the $2-10$~keV X-ray spectra 
of AGN is an iron emission line at energy 6.4~keV, corresponding to 
the Fe-K $n=2-1$ transition of ``cold'' (i.e., $\leq$ FeXVII) iron.
The line is usually ascribed to emission due to fluorescence in the 
inner part of the accretion disk. The typical EW of 
the line is $100-200$~eV. There is also evidence for a broad ``red wing''
extending to lower energies (Tanaka et al.\ 1995; Nandra et al.\ 1997). 
Once thought to be widespread, {\em XMM-Newton\/} spectra now show signs 
of this red wing only in a few objects (MCG-6-30-15 being the clearest
example, Vaughan \& Fabian 2004; see Fig.~\ref{risaliti_ironline}). 
This red wing has caused great excitement as a likely physical cause
is the gravitational redshift and relativistic Doppler shift of an 
Fe-K line originating from an accretion disk at only a few 
Schwarzschild radii ($R_S$) from the central black hole.
Such a broad line would be one of the best tools to look for general 
relativistic effects in strong gravity. Asymmetric profiles have been 
calculated for lines emitted at a few Schwarzschild radii
from non-rotating (Fabian et al.\ 1989) and rotating (Laor 1990)
black holes.

A second, narrow component of the Fe-K line is very clearly present 
in most AGN. The width of this ``narrow'' line (which is unresolved 
in CCD spectra from {\em ASCA\/}, {\em Chandra\/} ACIS, or 
{\em XMM-Newton\/} EPIC) is a few 1000~km~s$^{-1}$ or smaller 
when measured with the {\em Chandra\/} HETG spectrograph. 
This is similar to the width of optical 
and UV broad emission lines. This narrow component does not vary when 
the continuum varies, even for delay times of days. Coupled with 
the line width, this suggests an origin well beyond a few $R_S$, although a 
small radius is not fully ruled out (Fabian et al.\ 2002).
In the next few years, the {\em ASTRO-E2\/} satellite, with 6~eV 
resolution ($R=1000$) and an effective area of $\sim150$~cm$^2$, 
is expected to do much better in understanding this issue.
If the reflector is highly ionized, the peak energy of this line can 
be shifted toward high energies (6.7~keV for helium-like iron and
6.96~keV for hydrogen-like iron). It is also possible that two narrow 
components are present in the spectrum, one emitted by a cold reflector 
and the other by an ionized reflector. CCD detectors like the 
{\em ASCA\/} SIS or {\em XMM-Newton\/} EPIC (with energy resolution
of $\sim 120-150$~eV at 6~keV) are unable to separate these two 
lines, while the {\em Chandra\/} HETG spectrograph has limited effective 
area ($\sim40$~cm$^2$) at 6~keV.

\begin{figure}
\centerline{\includegraphics[width=12cm]{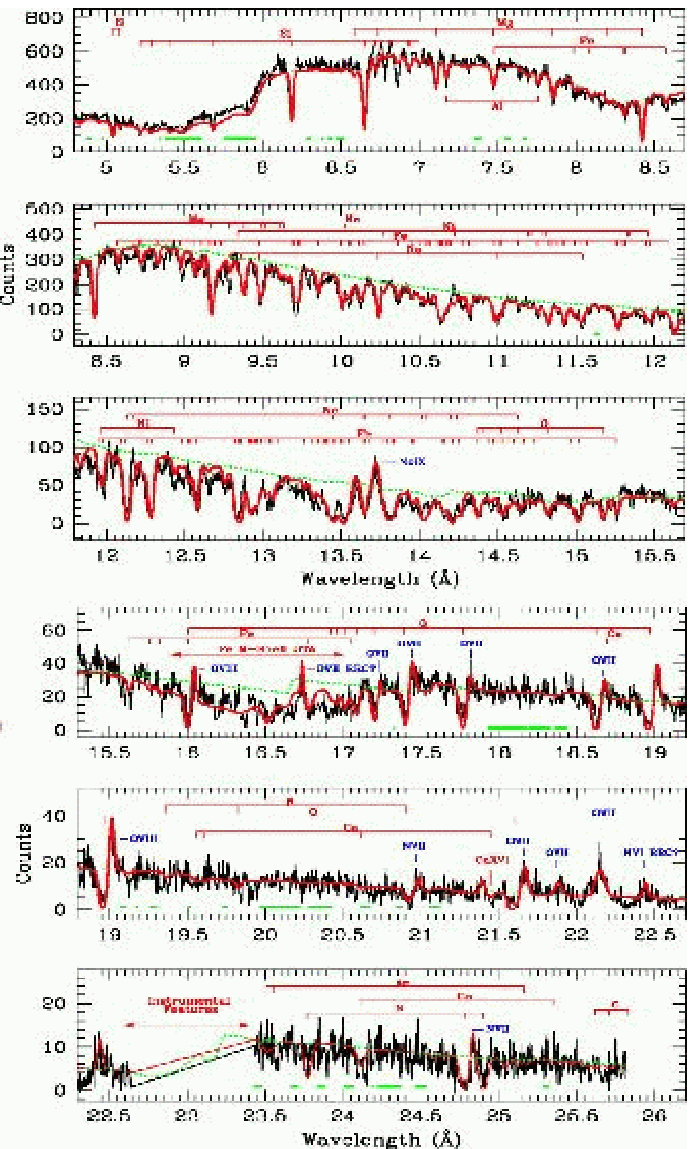}}
\caption{{\em Chandra\/} grating spectrum of NGC~3873, superimposed with
a simple two-component model fitting most of the absorption features. 
Figure is from Krongold et al.\ (2003; their Fig.~3).
}
\label{risaliti_grating}
\end{figure}

$\bullet$ {\em Warm absorbers.\/}
Warm absorber features are present in the soft X-ray spectra of half 
of the bright Seyfert 1 galaxies observed with {\em ASCA\/} 
(Reynolds et al.\ 2000). Recently, the availability of high resolution 
soft X-ray spectra, obtained with the grating instruments onboard 
{\em Chandra\/} and {\em XMM-Newton\/}, show that this component is formed 
by an outflowing gas. We show in Figure~\ref{risaliti_grating} the
highest signal-to-noise high resolution spectrum of an AGN, obtained
with a long observation of the Seyfert 1 galaxy NGC~3783. Recently, 
Krongold et al.\ (2003) were able to reproduce all of the observed lines 
with a two-phase absorber, with the two phases in pressure equilibrium.

%
%

\subsection{Bolometric Corrections} 
\label{risalitisecbolo}

Given an intrinsic dispersion in the SED of quasars, any flux-limited 
sample selected in a given spectral band is biased towards high ratios 
between the flux in the selection band and the bolometric emission.
This effect must be carefully taken into account in the construction 
of an AGN SED.

A well investigated example of the relevance of this selection effect 
is the average value of the optical-to-X-ray flux ratio
$\langle\alpha_{OX}\rangle$ obtained in different quasar samples. 
E94 estimate $\langle\alpha_{OX}\rangle=-1.35$ for their X-ray selected 
sample. On the other hand, Laor et al.\ (1997) find 
$\langle\alpha_{OX}\rangle=-1.55$ for local, optically selected PG quasars.
The difference, $\Delta\alpha_{OX}=0.2$, is a factor of $\sim 3.3$ in the
flux ratio. It is possible to use the distribution of observed 
$\alpha_{OX}$ in the two samples to estimate the effect of the selection 
bias. Elvis et al.\ (2002) showed that after this correction, the values 
estimated from the two samples match, with
$\langle\alpha_{OX}\rangle=-1.43$. 
These corrections are important when estimating the accretion 
luminosity of the universe and comparing this with the mass spectrum 
of local black holes (Fabian \& Iwasawa 1999; 
Elvis, Risaliti, \& Zamorani 2002).

%
%
\begin{figure}
\centerline{\includegraphics[width=12cm]{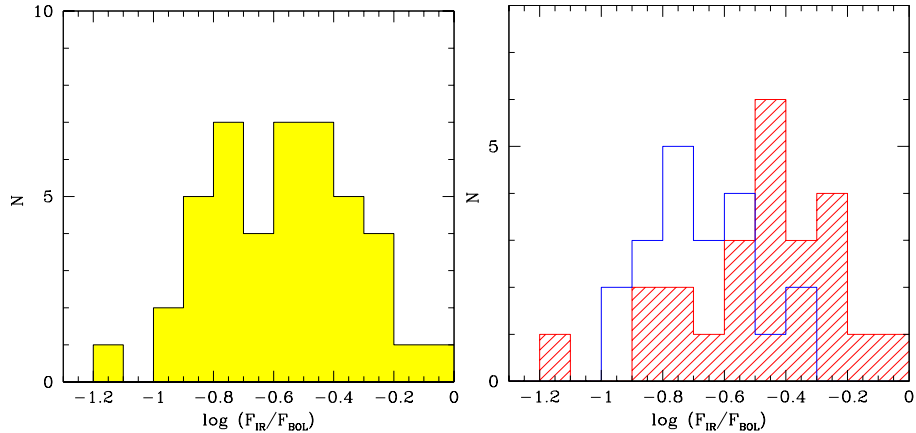}}
\caption{(a) Distribution of logarithmic IR-to-bolometric 
ratio for local ($z<0.4$) PG quasars observed with {\em ISO\/} 
(Haas et al.\ 2003). (b) Same, for objects with luminosity 
$L_{IR}<10^{12}$~ergs~s$^{-1}$ {\em (empty histogram)\/} and
$L_{IR}>10^{12}$~ergs~s$^{-1}$ {\em (shaded histogram)\/}.}
\label{risaliti_ir_isto}
\end{figure}

The analogous correction has not been computed so far for the IR
emission of PG quasars. We do this here. In Figure~\ref{risaliti_ir_isto}a,
we plot the logarithmic ratio $\alpha_{IR}$ of the IR
($3-1000\mu$m) to $>1\mu$m (bolometric) emission for the $z<0.4$ 
PG quasars observed with {\em ISO\/} (Haas et al.\ 2003). 
The emission from 2 to 100$\mu$m has been estimated from the $B$-band 
magnitude and the bolometric correction in E94. Approximating 
the distribution in Figure~\ref{risaliti_ir_isto}a with a Gaussian 
with a mean observed ratio $\langle\alpha_{IR}\rangle_{obs}=-0.56$ 
and $\sigma=0.3$, one obtains that the average $\alpha_{IR}$, 
corrected for the observational bias, is 
$\langle\alpha_{IR}\rangle$=$\langle\alpha_{IR}\rangle_{obs}+\sigma^2/2=-0.51$. 
The corresponding fraction of the bolometric luminosity emitted in the 
IR is 31\%. 

A summary of the average contribution of several spectral bands 
to the bolometric emission of local quasars is shown in 
Table~\ref{risaliti_bol_corr}. In this compilation, we made use of the 
data discussed above on PG quasars, as well as {\em HST\/} data of optically 
selected quasars (Telfer et al.\ 2002). These data do not show any 
spectral dependence with redshift in the optical/UV and therefore 
are assumed to be representative of local quasars. The $1-3\mu$m 
continuum, which is not covered in any of the works discussed above, 
has been taken from E94.

\begin{table}
\caption[Bolometric corrections for radio-quiet quasars, 
primarily $z<0.4$ PG quasars]
{Bolometric corrections for local quasars
}
\label{risaliti_bol_corr}
\centering
\begin{tabular}{lcccc}
\sphline
\em Band & \em Range ($\lambda$) & \em Range ($\nu$) & \em Range (energy) &\em F/F$_{TOT}$ \\
\sphline
Radio  & 3m-0.1mm          & 10$^8$-3$\times10^{11}$ Hz& 4$\times10^{-7}-1.2\times10^{-3}$ eV & 0\% \\
Submillimeter &1000-150$\mu$m     & 3-20$\times10^{11}$ Hz    & 1.2-8.3$\times10^{-3}$ eV & 0.2\% \\
far-IR    &150-40$\mu$m       & 2-7.5$\times10^{12}$ Hz   & 8.3-31$\times10^{-3}$ eV  & 4.9\%  \\
mid-IR &40-10$\mu$m        & 7.5-30$\times10^{12}$ Hz  & 3.1-12$\times10^{-2}$ eV  & 13.9\% \\
mid-IR &10-3$\mu$m         & 3-10$\times10^{13}$ Hz    & 0.12-0.41 eV              & 11.9\% \\
near-IR    &3-1$\mu$m          & 1-3$\times10^{13}$ Hz     & 0.41-1.25 eV              & 7.0\%  \\
Opt    &1$\mu$m-3000~\AA   & 3-10$\times10^{14}$ Hz    & 1.25-4.16 eV              & 12.2\% \\
UV     &3000-1200~\AA      & 1-2.5$\times10^{15}$ Hz   & 4.16-10.4 eV              & 16.5\% \\
EUV$^a$&1200-12.5~\AA      & 2.5-240$\times10^{15}$ Hz & 10.4 eV-1.0 keV           & 29.1\% \\
X-ray  &12.5-0.125~\AA     & 2.4-240$\times10^{17}$ Hz & 1-100 keV                 & 4.2\%  \\
\sphline
\end{tabular}
\begin{tablenotes}
Table notes --- {$^a$Based on high redshift quasars (Telfer et al.\ 2002).}
\end{tablenotes}
\end{table}

\subsection{Luminosity and Redshift Effects}

The SED described above is representative of local, optically selected 
quasars. Here we summarize the evidence for luminosity or redshift 
dependence in the emission of quasars.
The optical and UV spectra of quasars observed with {\em HST\/} and the 
SDSS show no evidence of a dependence on redshift or luminosity.
On the other hand, the optical-to-X-ray ratio $\alpha_{OX}$ shows 
clear evidence of a luminosity or redshift dependence in optically 
selected samples (Zamorani et al.\ 1981; Avni \& Tananbaum 1982, 1986;
Wilkes et al.\ 1994; Yuan et al.\ 1998; Bechtold et al.\ 2002; 
Vignali et al.\ 2003). Yuan et al.\ (1998b) discussed the reality of 
this effect in the {\em ROSAT\/} sample and concluded that the 
luminosity and/or redshift dependence could be due to selection 
effects, provided that the intrinsic dispersion in the X-ray emission 
of quasars is greater than in the optical. 

\begin{figure}[tbh]
\centerline{\includegraphics[width=10cm]{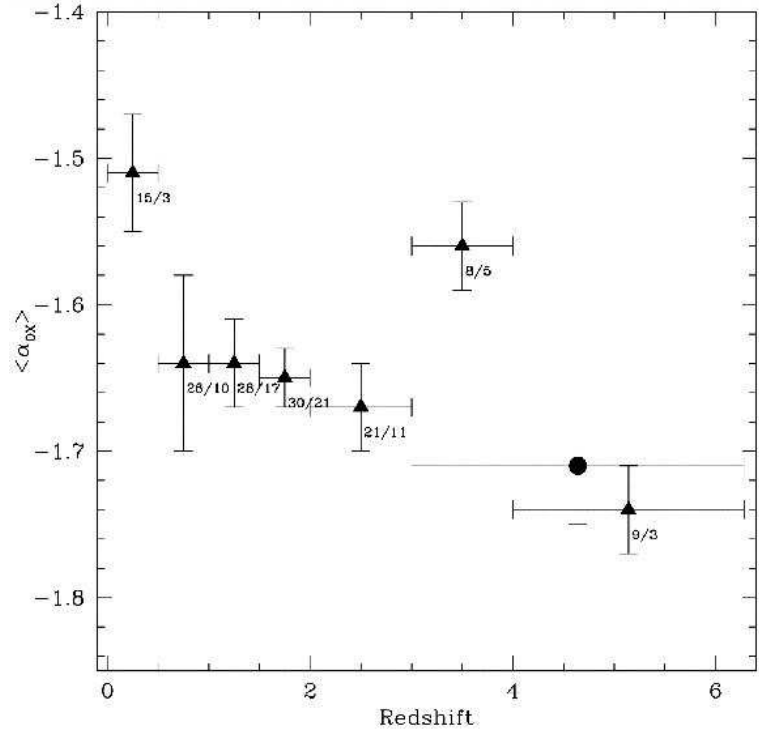}}
\caption{Dependence of $\alpha_{OX}$ on optical luminosity in
SDSS quasars observed with {\em ROSAT\/}.
Triangles are averages 
using the number of quasars indicated next to each point
({\em first number\/}---total number of quasars in the luminosity
interval; {\em second number\/}---number of X-ray upper limits 
in the luminosity interval). Circle is an average using all
quasars at redshifts z$>3$. Figure from
Vignali et al. (2003; their Fig.~7b).
}
\label{risaliti_aox}
\end{figure}

The latest results from the SDSS (Vignali et al.\ 2003), however, 
strengthen the observational evidence for a dependence. The statistical 
analyses performed on the SDSS quasars (Vignali et al.\ 2003)
and on the sample of optically selected quasars observed with 
{\em ROSAT\/} (Yuan et al.\ 1998) suggest that the dependence is only on
luminosity and not on redshift. However, it remains difficult
to disentangle the dependence on redshift and luminosity, which
 are strongly correlated in flux-limited samples.
In Figure~\ref{risaliti_aox}, we show
the $\alpha_{OX}$-luminosity correlation for a sample of SDSS quasars 
observed (mostly serendipitously) with {\em ROSAT\/} and {\em Chandra\/} 
(Vignali et al.\ 2003). The best-fit linear correlation is 
$\alpha_{OX}=-0.11\times\log L_\nu(2500$~\AA)+1.85.

%
%

The dependence of the IR emission of quasars on luminosity is much
harder to estimate, mainly because of the possible contribution from
star formation. In Figure~\ref{risaliti_ir_isto}b, we plot the 
distribution of the IR-to-bolometric ratio for the same sample 
as Figure~\ref{risaliti_ir_isto}a, but for 
two luminosity ranges: $L_{IR}<3\times10^{12}~L_\odot$
{\em (shaded histogram)\/} and $L_{IR}>3\times10^{12}~L_\odot$ 
{\em (open histogram)\/}.
Apparently, higher luminosity sources have, on average, a smaller 
fraction of their emission in the IR. Haas et al.\ (2003) concluded 
that most of the observed emission is due to the AGN. The same 
conclusion was reached by Kuraszkiewicz et al.\ (2003) for {\em ISO\/} 
SEDs of X-ray selected AGN. However, we 
cannot exclude the possibility that the effect in 
Figure~\ref{risaliti_ir_isto}b is due to a higher contamination 
by nuclear star formation in lower luminosity sources.

\subsection{Intrinsic Dispersion}
\label{risalitisubsecdisp}

\begin{figure}[tbh]
\centerline{\includegraphics[width=12cm]{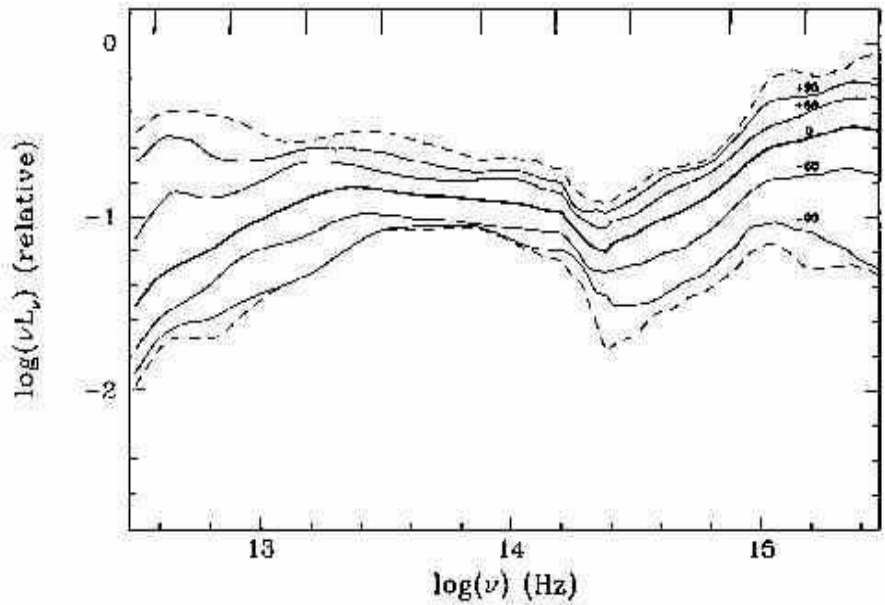}}
\caption{Intrinsic dispersion in the SED of E94.
Curves represent the 99\%, 90\%, and 68\% dispersion with respect
to the best fit SED and are normalized in order to have the same
flux at 1.25~$\mu$m. 
}
\label{risaliti_dispersion}
\end{figure}

As we will point out later, different quasar selection criteria 
produce different SEDs. Even with an homogeneous selection, the 
dispersion in the SEDs of AGN is rather large, about an order of
magnitude in the IR and UV, even when normalized at the 1$\mu$m 
``inflaction point''. The dispersion in the SED of local X-ray 
selected quasars is emphasized in E94 and shown in 
Figure~\ref{risaliti_dispersion}, but it is often not taken into 
account.

%
%

We believe this dispersion to be a fundamental property of quasars
that always should be considered when referring to an average SED. 
The reason is both ``physical'', in order to have
a physically correct view of quasar emission, and 
``computational'', since the results obtained when computing integral 
properties of quasar samples can be significantly altered by a 
non-zero dispersion distribution of parameters. An example is 
the correction from the observed to the ``effective'' $\alpha_{OX}$ 
discussed in \S\ref{risalitisecbolo}.
 
\begin{figure}[tbh]
\centerline{\includegraphics[width=11cm]{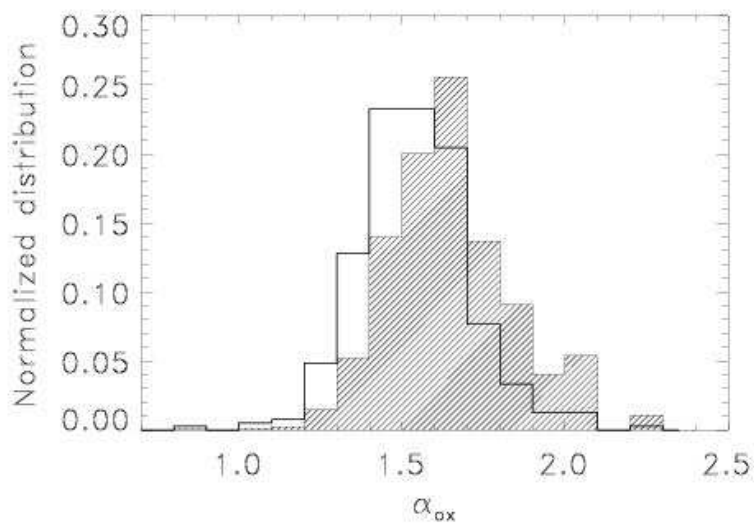}}
\caption{Distribution of $\alpha_{OX}$ for a sample of $\sim1000$ 
optically selected quasars observed with {\em ROSAT\/}. 
Empty histogram shows the {\em ROSAT\/} detections; shaded histogram 
takes into account upper limits, as well.
Figure from Yuan et al.\ (1998; their Fig.~9)
}
\label{risaliti_yuan}
\end{figure}

We already discussed the dispersion in the IR emission
in \S\ref{risalitisubsecdisp}. Here we only note that,
even if part of the observed dispersion 
(see Fig.~\ref{risaliti_ir_isto}) 
is due to a star formation contribution, it is likely that the
intrinsic dispersion in the IR-to-bolometric ratio is a factor
of $\sim 2$. Finally, in 
Figure~\ref{risaliti_yuan} we show the distribution of 
$\alpha_{OX}$ for the sample of optically selected quasars observed 
with {\em ROSAT\/} (Yuan et al.\ 1998). Approximating the distribution 
with a Gaussian, the standard deviation is $\sigma(\alpha_{OX})\sim0.2$,
corresponding to a dispersion in the ratio between optical and X-ray 
emission of a factor of $\sim3$. 

%
%

\section{SEDs of Obscured AGN}
\label{risalitisecseds}

Most of the AGN emission in the universe is obscured.
Locally, optically obscured (``type II'') AGN outnumber 
unobscured AGN (``type I'') by a factor of $\sim4$
(Maiolino \& Rieke 1995). This factor could be even higher at
redshifts $z=0.5-1$, according to synthesis models of the X-ray 
background (Gilli et al.\ 2001). Absorption introduces a further 
spread in the emission properties of AGN, since the observed 
SEDs depend both on the intrinsic emission (which has a large
dispersion) and on the amount, composition, and geometry of the 
absorber. In general, the main absorption mechanisms in AGN are
the following:\\

\noindent
(1) {\em Line absorption due to atomic transitions.\/} 
The strongest features observed in AGN spectra 
(``Narrow Absorption Lines'' and ``Broad Absorption 
Lines'' in the UV, ``Warm Absorption'' in the soft X-rays)
are mainly due to resonant absorption lines. \\

\noindent
(2) {\em Continuum absorption in the IR to UV due to dust.\/} \\

\noindent
(3) {\em Continuum absorption (or scattering) in the X-rays 
due to photoelectric absorption by dust and gas.\/} \\

The physical state (temperature, density), column density, and 
metallicity of the gas, the chemical composition of the dust, 
the dust-to-gas ratio, and the composition of the dust grains are 
all elements that affect the observed SED. Here we review the main 
emission properties of obscured AGN, focusing mostly on the actual 
observational results and only briefly discussing their physical 
interpretation.

\subsection{Radio/IR}

The radio emission properties of optically obscured AGN are similar 
to those of unobscured, type I AGN. A flat-spectrum, compact radio 
core is present in local AGN, with brightness temperatures 
$T_B>10^5$~K. Recent VLA observations (Nagar et al.\ 2000)
also revealed these radio cores in low luminosity 
($L<10^{41}$~ergs~s$^{-1}$) AGN, showing that radio emission is an 
ubiquitous property of all AGN. However, the fraction of luminosity 
emitted in the radio band is in all cases negligible with respect to 
the bolometric luminosity.

In some cases, free-free absorption can alter the observed radio 
spectrum of an AGN covered by a compact layer of warm gas 
(Neufeld et al.\ 1994). At high radio luminosity, HI absorption 
is also common (Veron-Cetty et al.\ 2000). HI and/or free-free 
absorption is preferentially found in X-ray heavily absorbed 
(column density $N_H>10^{24}$~cm$^{-2}$), optically type II
sources (Risaliti, Woltjer, \& Salvati 2003b).
 
\subsection{Near-IR/UV} 

The optical to UV band continuum emission is affected by dust absorption. 
Typically, the optical/UV continuum in type II AGN is heavily absorbed by 
dust and thermally reradiated at longer wavelengths. The observed continuum 
is dominated by the stellar contribution of the host galaxy and/or
scattered emission. This latter component can be disentangled through 
observations in polarized light (see below for further details). 

\begin{figure}[hbt]
\vskip.2in
\centerline{\includegraphics[width=10cm]{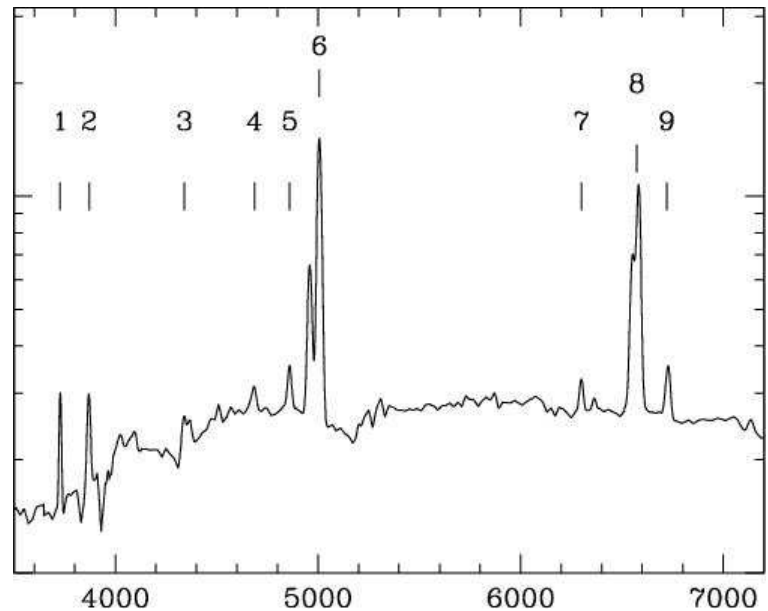}}
\caption{Optical spectrum of the prototype Seyfert 2 galaxy NGC~1068. 
Main emission lines are 1: [OII]$\lambda$3727\AA, 
2: [NeIII]$\lambda$3869\AA, 3: H$\gamma$, 4: HeII $\lambda$4687\AA, 
5: H$\beta$, 6: [OIII]$\lambda$5007\AA, 7: [OI]$\lambda$6300\AA,
8: H$\alpha+$[NII]$\lambda$6585\AA, 9: [SII]$\lambda$6732\AA.}
\label{risaliti_spesey2}
\end{figure}

The emission-line spectrum is dominated by ``narrow'' emission lines 
(typical widths $300-800$~km~s$^{-1}$) corresponding to forbidden 
atomic transitions plus Balmer lines. 
This implies that the emitting gas is located farther 
from the center than the broad emission line emitter (if the width is 
interpreted as Keplerian motion) and that its density is lower than 
$\sim10^5$~cm$^{-3}$. The broad emission lines are absent in objects 
classified as ``pure'' type II AGN (except for a weak scattered component 
visible in polarized light). Less obscured AGN are classified as type 1.9 
and type 1.8 and show broad components in, respectively, the H$\alpha$ 
line and both the H$\alpha$ and H$\beta$ lines. The optical spectrum of 
the prototype Seyfert 2 NGC~1068 is shown in Figure~\ref{risaliti_spesey2}.

%
%

In heavily obscured AGN, when only emission lines are seen, it can be
hard to distinguish an AGN from a starburst (this will be discussed in
\S\ref{risalitisectype2}).
A major indicator of the presence of an AGN is the high ratio of
high ionization lines, such as [OIII]$\lambda5007$~\AA\ or N~V, with
respect to low ionization lines, such as H$\beta$ or H$\alpha$.
These narrow emission lines are the result of the reprocessing of 
nuclear radiation by gas not covered by the nuclear absorber because 
it lies outside the obscuring region. The central emission (continuum 
and broad lines) can be scattered by circumnuclear hot gas, or by dust
from some region outide of the obscuring region that lies in a direction 
with a clear view of the central source. This component is typically 
too weak with respect to the galaxy emission to be seen in the total 
spectrum, but it clearly emerges in polarimetric observations. The 
implication is that the obscuring region has a flattened distribution.

Historically, the observation of broad lines in the polarized spectrum 
of NGC~1068 was fundamental for the formulation of the unified model of 
AGN, which states that type I and type II AGN are intrinsically the 
same objects and differ only in the orientation of the circumnuclear 
absorber (Antonucci \& Miller 1985).

\subsection{X-rays}

Obscuration in the X-rays is due to photoelectric absorption (dominant
below $\sim3$~keV) and Compton scattering (dominant from $\sim7$ to 
$\sim30$~keV). The X-ray spectral properties of obscured AGN depend 
on the amount of absorbing column density: column densities below 
$\sim1.5\times10^{24}$~cm$^{-2}$ produce a photoelectric cut-off at 
energies between 1 and 10~keV (in this case, the source is ``Compton thin'');
column densities between $\sim10^{24}$~cm$^{-2}$ and $\sim10^{25}$~cm$^{-2}$
absorb the X-ray primary emission up to several tens of keV; even higher 
column densities completely obscure the central source in the X-rays.

In heavily absorbed sources (``Compton thick'', $N_H>10^{24}$~cm$^{-2}$), 
the two main spectral features are a prominent
iron K$\alpha$ emission line with EW$\sim1-3$~keV,
and a reflected and/or scattered continuum.
In less obscured sources, the EW of the iron line depends on the fraction
of the intrinsic continuum emission absorbed at the line energy;
for $N_H<10^{23}$~cm$^{-2}$, values typical of type I AGN are 
observed (EW$\sim100-300$~eV), in agreement with the unified model.
The reflected/scattered component is the same as described in 
\S\ref{risalitisecxrays} for type I AGN. 

\begin{figure}[tbh]
\vskip.2in
\centerline{\includegraphics[width=12cm]{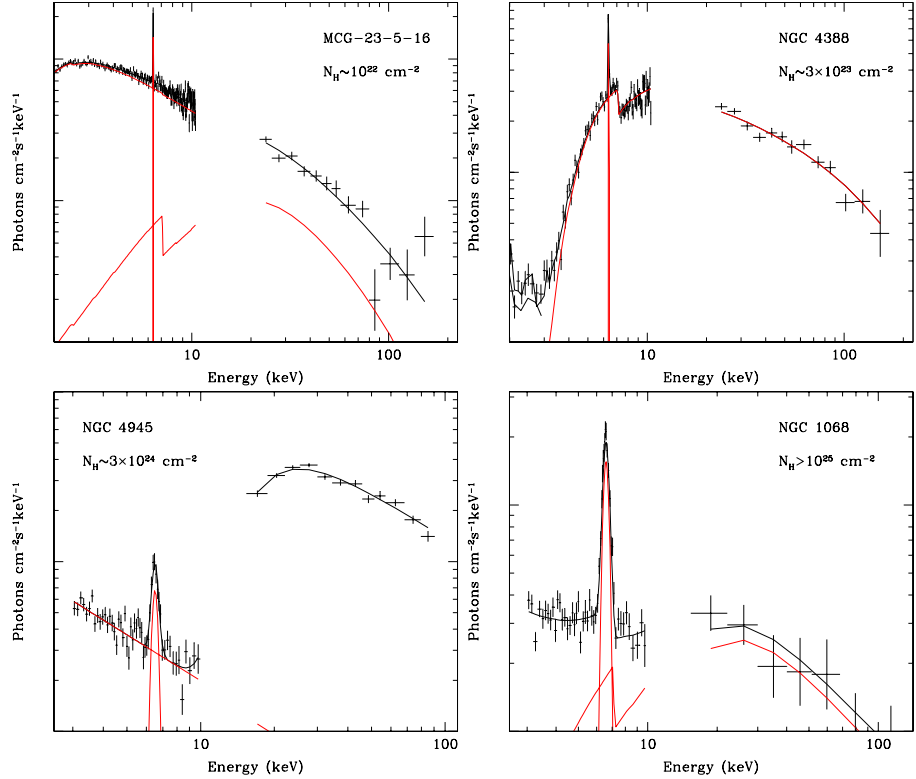}}
\caption{Four $2-100$~keV {\em BeppoSAX\/} best fit X-ray spectra
of Seyfert 2 galaxies. Main components of the best fit models are
also shown. MCG-5-23-16 and NGC~4388 (Risaliti 2002)
are ``Compton thin'', i.e., they are dominated by the primary emission
down to a few keV. In MCG-5-23-16, a cold reflection component also
gives a measurable contribution. The continuum in the Compton-thick
source NGC~4945 (Guainazzi et al.\ 2000) is due to a warm reflection
component in the $2-10$~keV range, while at higher energies the
intrinsic component emerges. Note the high ratio between the $10-100$~keV 
and the $2-10$~keV emission, as compared with the Compton-thin
sources. NGC~1068, also Compton-thick (Matt et al.\ 1999), shows a cold
reflection and a warm reflection component. Equivalent widths of the 
iron line are $\sim100$~eV in MCG-5-23-16, $\sim500$~eV in NGC~4388, 
and $1-2$~keV in NGC~4945 and NGC~1068.
}
\label{risaliti_xrays}
\end{figure}

In Figure~\ref{risaliti_xrays}, we plot the $1-100$~keV
spectra of four representative obscured AGN: 
MCG-5-23-16 ($N_H=10^{22}$~cm$^{-2}$, Risaliti 2002),
NGC~4388 ($N_H=4\times10^{23}$~cm$^{-2}$, Risaliti 2002),
NGC~4945 ($N_H=2\times10^{24}$~cm$^{-2}$, Vignati et al.\ 1999), and
NGC~1068 ($N_H>10^{25}$~cm$^{-2}$, Matt et al.\ 1999).

%
%

\subsection{Relation between Dust and Gas Absorption}
\label{secrisalitireln}

According to the standard paradigm of the unified model, X-ray 
absorption should be observed in optically type II objects. This 
is indeed what has been found in many local AGN. However, recent 
observations challenge this simple view and suggest a more 
complex scenario.

A direct measurement of the dust-to-gas ratio is possible in
objects with intermediate optical classification, i.e., 
Seyfert 1.8 or 1.9. These show clear signs of absorption in 
the optical/near-IR but still have broad components
in some of the brightest emission lines. In these cases, an 
estimate of the dust absorption can be obtained from the ratio of 
the emission line fluxes (typically, the hydrogen lines) and can be
compared with the absorbing column density measured in the X-rays.

For a Galactic dust composition and dust-to-gas ratio, the relation 
between optical extinction and X-ray absorption is 
$A_V\sim4.5\times10^{-21}~N_H$. Therefore, one would expect objects 
with optical broad lines to have X-ray column densities not higher 
than 10$^{22}$~cm$^{-2}$. Maccacaro, Perola, \& Elvis (1982) and 
Maiolino et al.\ (2001) analyzed a sample of bright, intermediate 
Seyferts and found that X-ray absorption is systematically higher 
than expected from optical extinction by a factor of $\sim10$. 
The physical explanation of this result may be a lower than 
Galactic dust-to-gas ratio, or a different composition of dust 
grains. The observed SEDs of these objects can be significantly 
different than standard type I and type II templates, with lower 
UV, larger IR, and heavily absorbed X-rays (Ward et al.\ 1982).

Moving to more extreme cases, BAL quasars 
are objects with no or little dust extinction in the optical/UV, 
but with broad, blueshifted, and often saturated absorption lines.
Many of them also only have extremely faint X-ray emission 
(note, however, that there are exceptions, such as the BAL 
quasars found in deep X-ray surveys; Barger et al.\ 2002).

\begin{figure}[hbt]
\vskip.2in
\centerline{\includegraphics[width=11cm]{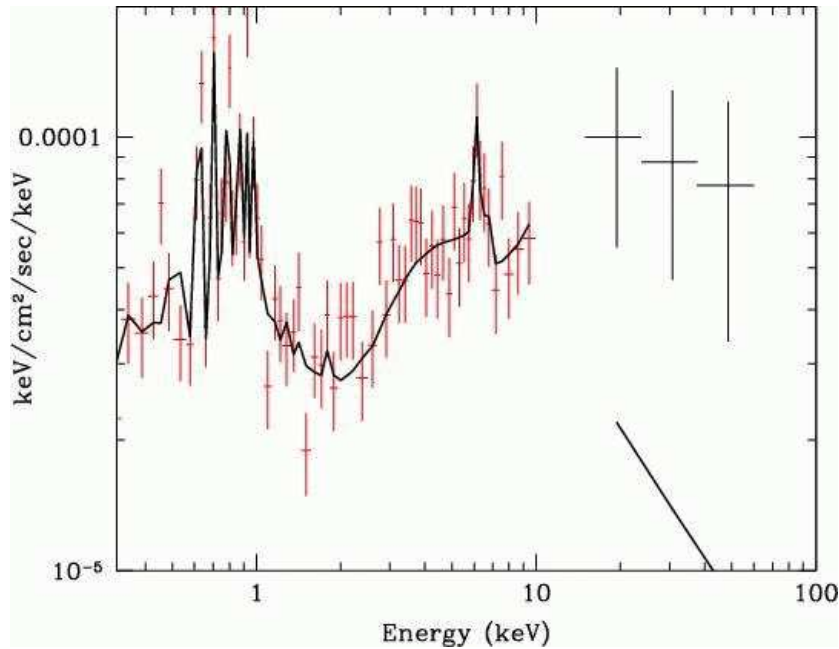}}
\caption{
{\em XMM-Newton\/} and {\em BeppoSAX\/} spectrum of MKN~231.
Figure obtained using data and model of Braito et al.\ (2004).
Model shown here is a best fit to the $0.5-10$~keV emission.
The large excess at E$>$10~keV is due to the intrinsic emission
of the AGN, which is absorbed by a column density $N_H\sim2\times
10^{24}$~cm$^{-2}$.
}
\label{risaliti_mkn231}
\end{figure}

There is now convincing evidence 
that these objects are intrinsically normal quasars, covered by a 
high column density of dust-free gas that is responsible for heavy 
absorption in the X-rays and absorption lines in the optical/UV.
A strong correlation between these two absorption features has been 
found by Brandt et al.\ (2001), and recent X-ray observations are 
starting to directly measure the X-ray column densities of these objects. 
A clear example is the {\em BeppoSAX\/} observation of the X-ray weak 
BAL quasar MKN~231, which revealed powerful hard X-ray emission 
above 10~keV obscured by a column density $N_H>10^{24}$~cm$^{-2}$
(Braito et al.\ 2004, see Fig.~\ref{risaliti_mkn231}).

%
%

Other examples of objects with strongly differing optical and X-ray 
absorption are found in quasar surveys where selection criteria
other than optical/UV color are used. An interesting example is the 
sample of ``red'' quasars discovered with the 2MASS near-IR survey 
(Cutri et al.\ 2001). The selection criterion adopted in this case
is $J-K>2$, which is efficient at low redshifts where the minimum 
in the quasar emission, due to the sublimation temperature of dust,
is observed in the $J$ band. {\em Chandra\/} observations 
of a sample of these ``red'', yet broad-line objects 
(Wilkes et al.\ 2002), revealed that they are extremely faint in the 
X-rays, probably due to absorption by a column density of order 
10$^{23}$~cm$^{-2}$. 

Another case of X-ray weakness in optically 
type I quasars is found in the sample of quasars from the Hamburg 
survey (Hagen et al.\ 2001) which also have {\em ROSAT\/} observations 
(Risaliti et al.\ 2001). Most of these objects, which are slightly 
redder in the optical than standard blue quasars (the selection 
criterion was based both on blue color and on low resolution 
spectroscopy), are undetected by {\it ROSAT}, contrary to what was
expected assuming a ``normal'' $\alpha_{OX}$. Subsequent {\em Chandra\/} 
observations of a subsample of these objects revealed that they are 
underluminous in the X-rays by a factor of $\sim3$ to $\sim100$ with 
respect to PG quasars (Risaliti et al.\ 2003a). It is not clear whether 
the observed X-ray weakness is an intrinsic property of these objects 
(as most of the {\em Chandra\/} spectra seem to suggest) or whether it
is due to absorption. In any case, they are part of a population of 
quasars with SEDs different than that of standard type I or type II AGN.

In addition to the cases described above of optically type I AGN with 
the X-ray properties of type II AGN, several examples are known of the 
opposite case, i.e., objects with type II optical properties and no 
hint of absorption in the X-rays. Panessa et al.\ (2003) described a 
sample of local sources optically classified as Seyfert 2 galaxies
with no measured X-ray absorbing column density in excess of the 
Galactic value.

\section{Finding Obscured AGN in the Universe}
\label{risalitisectype2}

In the previous two sections, we described the main properties of AGN 
continua for both obscured and unobscured sources, while neglecting 
the problem of disentangling the AGN emission from that of the host 
galaxy. Type I AGN are easily detected in the optical down to 
luminosities that are intrinsically weaker than the total host galaxy 
emission. In this case, high signal-to-noise and careful galaxy 
subtraction can detect the central AGN emission 
(Ho, Filippenko, \& Sargent 1999). Type II AGN, however, can be 
extremely elusive, even when they dominate the bolometric emission 
of the galaxy, since most of their primary emission is thermally 
reradiated into the IR. Since the IR is simply a sum of blackbodies, 
to a good approximation all signatures of the origin
of the luminosity are lost. Hence it is extremely difficult to 
distinguish an AGN contribution from that of star-forming regions. 
This problem is particularly important in the study of high 
luminosity sources, such as the Ultraluminous Infrared Galaxies 
(ULIRGs; Sanders \& Mirabel 1996) and their probable high redshift 
analogs, the powerful submillimeter emitters detected in SCUBA surveys
(e.g., Smail, Ivison, \& Blain 1997; Barger et al.\ 1998; 
Hughes et al.\ 1998). 
Distinguishing AGN contributions from star formation is of great 
importance in determining which class of source makes up the 
submillimeter/IR background, and therefore in knowing the relative 
contributions that accretion onto supermassive black holes and 
star formation make to the total luminosity of the universe.

\subsection{Indicators for (Local) AGN and Starbursts}

We next discuss the main indicators of optically obscured AGN 
(hereafter, we use ``obscured'' to mean that the optical/UV continuum 
and broad emission line spectrum is not observable) in each wavelength 
band, and the limits of each technique. Then we try to draw 
some general conclusions on the possibility of detecting AGN activity
in galaxies, now and in the foreseeable future.
Two main elements are relevant for the effectiveness of the different
indicators: (1) the amount of X-ray absorbing column density $N_H$ 
and optical extinction $A_V$, and (2) the fraction of solid angle 
covered by the obscuring medium.

$\bullet$ {\em X-rays.\/} If the X-ray absorbing column density is not 
larger than 10$^{25}$~cm$^{-2}$, the direct $>10$~keV X-ray emission of 
the AGN can penetrate the absorber. In this case, the hard X-ray 
emission will be at least an order of magnitude higher than
that of the host galaxy, down to luminosities $\sim10^{41}$~ergs~s$^{-1}$.
With arcsecond {\em (Chandra)\/} resolution, the contrast is improved by a 
factor of $10-100$ for $z\leq 0.1$, though {\em Chandra\/}'s effective 
upper energy bound of $\sim7$~keV limits the detection to 
$N_H<10^{23}$~cm$^{-2}$ (this limit moves up to 10$^{24}$~cm$^{-2}$ for
objects at redshifts $z\sim$1). In these cases, the detection of the AGN
is unambiguous. 

The contribution of the host galaxy to the hard 
X-ray emission is usually modeled by two components: the thermal 
emission due to warm interstellar gas (kT$\sim$0.1-1~keV),
and the contribution from compact sources (dominated by X-ray binaries)
that, on average, is reproduced by a power law with $\alpha\sim-0.5$ to 
$-0.7$, or by a thermal component with kT$\sim$20~keV (Fabbiano 1989).
The luminosity of these components can be of the same order 
as---or greater than---that of the obscured AGN 
($L_X<10^{40}$~ergs~s$^{-1}$ for spirals;
$L_X\sim10^{41}$~ergs~s$^{-1}$ for ellipticals and starbursts). 
From the observed direct X-ray emission, it is possible to give 
a rough estimate to the bolometric emission of the AGN using 
the information of the average SED discussed in \S\ref{risalitisectype1}. 
Examples of the effectiveness of hard X-ray observations include
ULIRG MKN~231, shown in Figure~\ref{risaliti_mkn231} (discussed 
in \S\ref{secrisalitireln} as a BAL quasar), and NGC~6240,
an ULIRG with no evidence of AGN activity below $20-30$~keV 
(except, perhaps, for some indication from mid-IR coronal lines, 
Lutz et al.\ 2003). A {\em BeppoSAX\/} observation of NGC~6240 discovered 
a powerful AGN with a column density $N_H>10^{24}$~cm$^{-2}$ that 
showed up at energies $E>20$~keV (Vignati et al.\ 1999). 

However, despite these impressive examples, this method has so far 
proved to be useful only in a few cases. A {\em BeppoSAX\/} search for 
ULIRGs known to host an AGN from other indicators (described below) 
in the $10-100$~keV band failed to detect the nuclear activity in 
most cases (Risaliti et al.\ 2004, in preparation). This implies
that most of these sources are obscured by column densities 
$N_H>10^{25}$~cm$^{-2}$.

If $N_H>10^{25}$~cm$^{-2}$, no direct emission can penetrate the 
obscuring medium because multiple Compton scatterings gradually 
remove energy from the photons until they can be photoelectrically 
absorbed. In this case, the only way to detect the AGN is through 
reflected (scattered) emission. The main spectral properties of a 
cold reflection dominated AGN are a flat spectrum ($\alpha>-1$ in 
the $2-10$~keV band) and a prominent iron line (EW$>$1~keV)
at $\sim6.4$~keV. In principle, two methods can be used to disentangle
the two contributions, based on spatial and spectral analyses, 
respectively. \\

\noindent
1) High spatial resolution can help to resolve the nuclear 
region where the AGN emission is dominant. This approach is obviously most 
useful when used on nearby sources. A case study, which shows both 
the power and the limitations of this approach, is the {\em Chandra\/} 
observation of the ULIRG Arp~220. The subarcsecond resolution instruments 
of {\em Chandra\/} resolved a weak point-like hard X-ray source
with luminosity $\sim4\times10^{40}$~ergs~s$^{-1}$ (Clements et al.\ 2003),
which is probably associated with an AGN. No previous spectral analysis 
was able to disentangle this component from the diffuse X-ray emission 
because of  the orders of magnitude worse spatial resolution. However, 
even in the Arp~220 case, we cannot rule out that the observed hard 
emission is due to $\sim10-100$ X-ray binaries in a region of intense
star formation (the size limit for a {\em Chandra\/} point source at 
the redshift of Arp~220, $z=0.018$, is $\sim400$~pc). \\

\noindent
2) Spectral decomposition of the AGN and starburst components 
requires high effective area. Recently, an {\em XMM-Newton\/} survey 
of nearby, bright ULIRGs provided the best X-ray spectra of ULIRGs 
(Franceschini et al.\ 2003). The AGN component was clearly detected 
in 3 out of 8 sources through the high EW 6.4~keV iron line
and the flat continuum. However, in the remaining five sources, 
the case is ambiguous, as the presence of a completely obscured 
($N_H>10^{24}$~cm$^{-2}$) AGN cannot be ruled out. \\

$\bullet$ {\em Optical/UV.\/} 
By definition, optically obscured (type II) 
AGN do not show any intrinsic continuum emission in the optical/UV 
band. The two main ways to search for AGN in these wavebands
are through scattered (polarized) light and narrow emission line
ratios. Both require that there are some unobscured directions 
out of the nucleus.

Scattering by warm electrons has been briefly discussed in 
\S\ref{risalitisecseds}. In the polarized spectrum, broad 
emission lines can be detected, unambiguously revealing a 
central AGN. 

%
%
\begin{figure}
\vskip.2in
\centerline{\includegraphics[width=9cm]{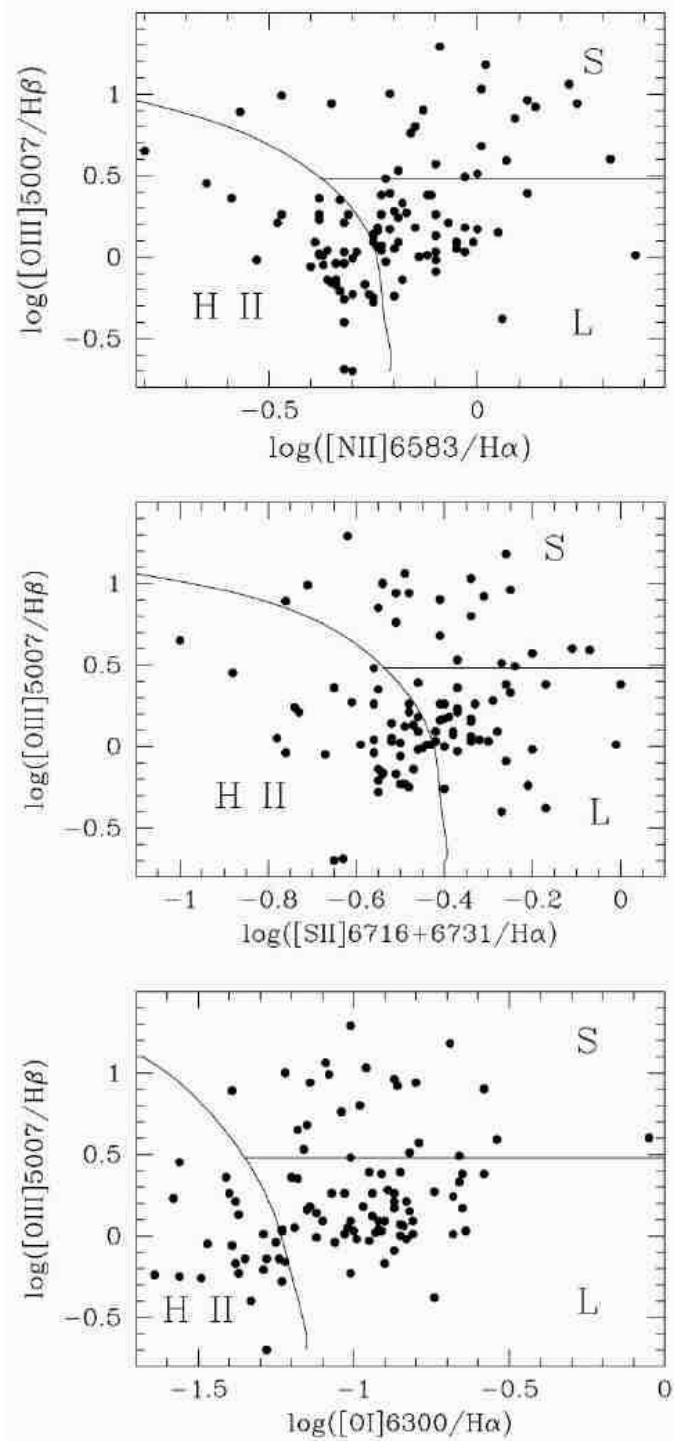}}
\caption{Diagnostic diagrams based on narrow emission line 
ratios for a sample of bright ultraluminous infrared galaxies. 
Figure from Veilleux, Kim, \& Sanders (1999; their Fig.~2).}
\label{risaliti_veilleux}
\end{figure}

The main method for classifying narrow emission line objects is 
through the ratio of emission line fluxes: high ionization lines
are expected to be stronger in AGN than in starbursts. 
The main optical emission lines shown in 
Figure~\ref{risaliti_spesey2}, and listed in the caption,
are roughly the same as in a spectrum of a starburst galaxy. 
In Figure~\ref{risaliti_veilleux}, we show a classical diagnostic 
diagram, first introduced by Veilleux \& Osterbrock (1987),
in which type II AGN are clearly separated from starbursts. In the 
same figure, a third class of objects is shown, the so-called 
LINERs (Low Ionization Emission Line Objects). The origin of their 
emission (nuclear activity or star formation) is not yet clear. 

$\bullet$ {\em Infrared.\/} 
A general indicator of AGN activity is a warm IR spectrum. 
The IRAS $25-60\mu$m color is rather effective in finding 
AGN-powered IR sources (see, for example, de Grijp et al.\ 1997).
However, while a high $25-60\mu$m color can be considered 
an indicator of AGN activity, independent confirmation is needed.

{\em ISO\/} has enhanced the analytic capability of IR analysis. 
For example, Laurent et al.\ (2000) have shown that the ratio 
between the $12-18\mu$m and the $5-8.5\mu$m emission (these are 
the ranges of two filters in the ISOPHOT instrument on the 
{\em ISO\/} satellite) is effective in discriminating the AGN and 
starburst contributions at redshifts between 0.4 and 1. At a finer 
level of detail, when low resolution spectroscopy is available, 
broad emission features in the mid-IR band can be used as effective 
AGN indicators. For example, the absence of a broad $7.7\mu$m
emission, due to Polycyclic Aromatic Hydrocarbon (PAH) molecules, 
is believed to be an indicator of AGN emission, both because of 
the strength of this indicator in control samples and for physical 
considerations: PAH molecules should be destroyed by the high energy 
continuum of an AGN.

%
%
\begin{figure}
\vskip.2in
\centerline{\includegraphics[width=6.7cm,angle=-90]{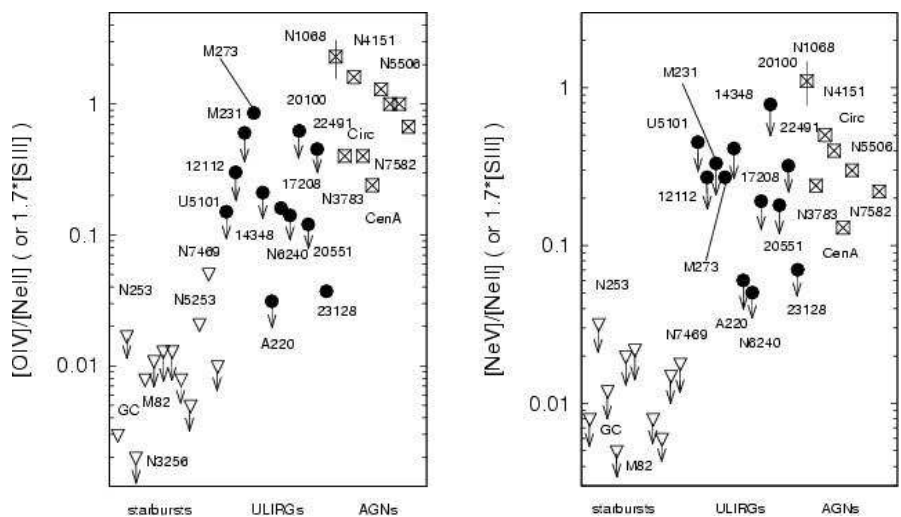}}
\caption{Diagnostic diagrams based on mid-IR emission lines.
Inverted triangles and squares denote sources optically classified as
starbursts and AGN, respectively. Circles denote ULIRGs.
Figure from Genzel et al.\ (1998; their Fig.~3). 
}
\label{risaliti_genzel}
\end{figure}

Finally, higher resolution {\em ISO\/} SWS spectra provide useful 
indicators of the relative AGN/starburst contributions. As in the 
case of optical lines, high excitation emission lines are tracers 
of AGN activity, while strong, low excitation lines are a sign of 
starbursting activity. For example, [O~IV]$\lambda25.9\mu$m and 
[Ne~V]$\lambda14.3\mu$m are among the strongest high excitation 
mid-IR emission lines, while [Ne~II]$\lambda12.9\mu$m and 
[S~III]$\lambda18.7\mu$m are strong low excitation lines. An example 
of the use of these indicators can be found in Genzel et al.\ (1998), 
where the [O~IV]/[Ne~II] and [Ne~V]/[Ne~II] ratios of a sample of 
nearby, bright ULIRGs are compared with those of comparison samples 
of local AGN and starbursts (see Fig.~\ref{risaliti_genzel}).
The power of this diagnostic is also well illustrated by its application
to the {\em ISO\/} spectrum of the galaxy NGC~6240. This source, as 
discussed above, is known from hard X-ray ($10-100$~keV) observations 
to host a powerful AGN, but it does not reveal any indication of an AGN 
at longer wavelengths, except for a strong [O~IV]$\lambda25.9\mu$m 
in the {\em ISO\/} SWS spectrum (Lutz et al.\ 2003). All the IR methods 
discussed above will be much more effective, and applicable at higher 
redshift and/or fainter objects, with the new observations of the 
{\em Spitzer Space Telescope\/}.

\begin{figure}
\vskip.2in
\centerline{\includegraphics[width=10cm]{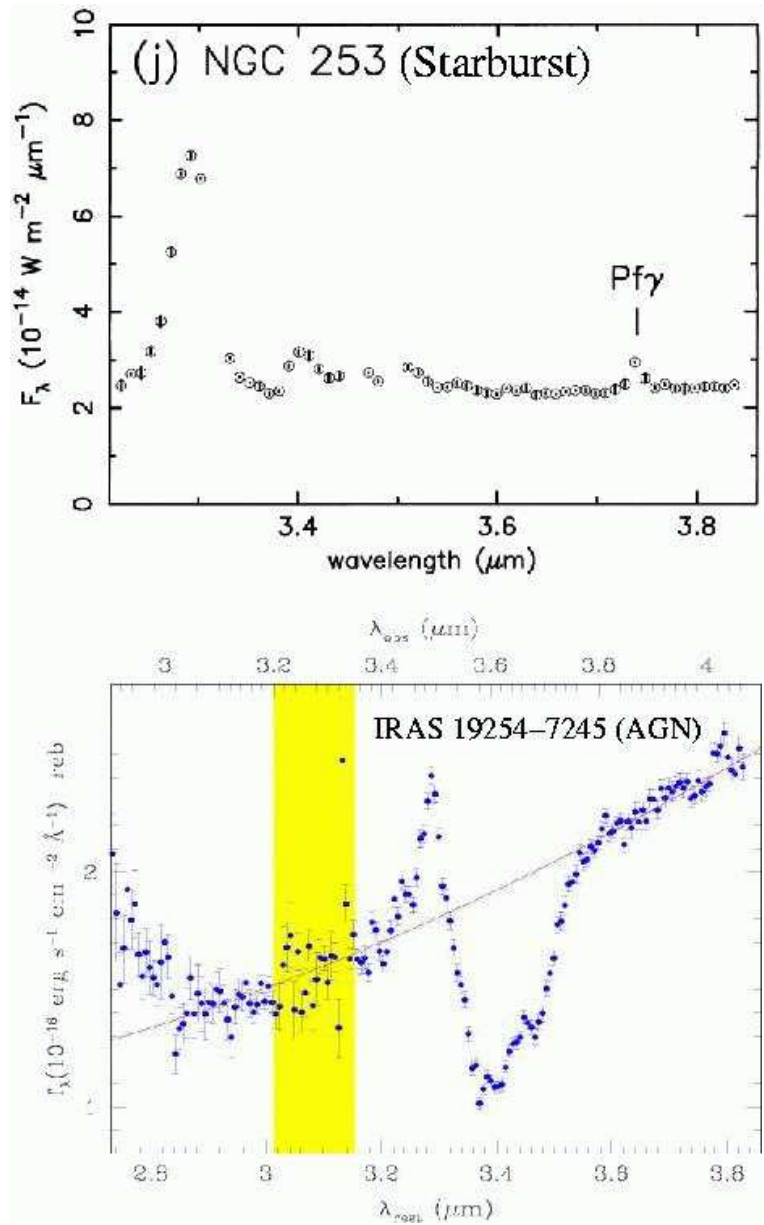}}
\caption{$L$-band spectrum of the starburst galaxy NGC~253 
(from Imanishi \& Dudley 2000; their Fig.~2j) and of the 
AGN-dominated ULIRG IRAS~19254+7245 (Risaliti et al.\ 2003c;
their Fig.~1). The
most evident features are the $3.3\mu$m PAH emission line in 
both spectra and the 3.4~$\mu$m absorption feature in 
IRAS~19254+7245. Shaded band in the spectrum of IRAS~19254+7245 
indicates a wavelength interval with low atmospheric transmission. 
\label{risaliti_ulirgs}
}
\end{figure}

Moving to shorter wavelengths, recent observations in the $L$-band 
from ground-based telescopes have provided interesting new ways to 
disentangle AGN and starburst activity: the $3.3\mu$m PAH emission 
feature is a starburst indicator (Tokunaga et al.\ 1991), 
AGN are characterized by an absorption feature at $\sim3.4\mu$m due 
to hydrocarbon dust grains (Pendleton et al.\ 1994), and the $3-4\mu$m 
continuum is much steeper in AGN-dominated sources. As examples, 
we plot in Figure~\ref{risaliti_ulirgs} a starburst 
spectrum (NGC~253, Imanishi \& Dudley 2000) and an AGN spectrum 
(IRAS~19254+7245, Risaliti et al.\ 2003c).

%
%

$\bullet$ {\it Radio.} 
High spatial resolution radio observations can provide useful 
indicators of AGN activity. Very Long Baseline Interferometry (VLBI) 
observations of ULIRGs
(Lonsdale et al.\ 1993; Smith et al.\ 1998) showed that compact, 
high brightness temperature (well above starburst temperatures;
T$_b>10^5$~K) sources are preferentally found in objects 
classified optically as AGN. Physically, compact and high 
luminosity radio emission can only be explained by AGN activity, 
or by assuming the simultaneous presence of several extremely 
luminous radio supernovae in regions with a radius of less than 
1~kpc, a highly unlikely event, even in the most intense starburst 
regions.

Recently, a radio VLA survey of the ULIRGs in the IRAS 1~Jy sample
(Kim \& Sanders 1998) showed that compact radio emission is present
in most LINER classified ULIRGs, similar to that found in objects
optically classified as AGN, and contrary to what is found in starburst
galaxies (Nagar et al.\ 2003). This result is particularly interesting 
since it provides a method to estimate the AGN incidence in a class of 
objects---LINERs---for which optical spectroscopic criteria do not 
give clear indications (as shown in Fig.~\ref{risaliti_veilleux}).

\subsection{Limitations}

At present, the methods described above can be applied only
to bright and local AGN (with a few exceptions, like the
hard X-ray diagnostics, and the ISOPHOT colors). 
Indeed, almost all of the examples shown here are about 
observations of nearby, powerful IR galaxies.
These sources are not only interesting in themselves but 
represent analogs of the population dominating the
submillimeter background, which probably comprises high redshift 
galaxies. Understanding the energy source of these populations 
would ultimately give us the ability to estimate the total 
fraction of the radiated energy of the universe due to AGN.
It is therefore very important to understand the perspective 
of extending these methods to fainter and more distant sources.

There are two main limitations of the methods described above,
one due to physical limits, and one due to observational 
capabilities. Observations of better quality can be done with
forthcoming instruments and can extend these diagnostics to 
a much wider range of sources. However, all the methods described 
above are based on the detection of either direct AGN emission
or scattered, reflected, or reprocessed (in the case of emission lines) 
emission. If the gas column density is higher than 
$\sim10^{25}$~cm$^{-2}$, the dust extinction is higher than 
$\sim8-10$ magnitudes (even in the IR), and the obscuration of the 
source is complete along all lines-of-sight, then none of the above 
methods can be used. Even the mid-IR emission will be self-absorbed and
reprocessed at longer wavelengths, so no big differences are expected 
in the far-IR spectra of AGN and starburst dominated sources.
In this extreme case, all the observed emission of the active 
nucleus will be thermally reprocessed by the circumnuclear obscurer. 
There is then only one method to search for AGN in these sources: 
spatial resolution. In principle, high spatial resolution is rather 
simple: if high far-IR emission is detected in too small of a region, 
so that even the most compact starburst can be ruled out, then only 
an AGN can be the energy source. In order to be effective, this 
requires resolutions of at most a few tens of parsecs for ULIRGs, 
which is far from being reached for the nearest ULIRGS, even with 
the new {\em Spitzer Space Telescope\/}. (One milliarcsec at $z=0.05$ 
corresponds to $\sim1$~pc with $H_0=70$~km~s$^{-1}$~Mpc$^{-1}$). 
However, this remains a possibility (maybe the only one) for a 
future solution to this problem.

In summary, will we ever be able to understand the origin of the
emission in the powerful far-IR and submillimeter sources and to
estimate the total contribution of accretion to the energy output
of the universe? The answer depends both on the improvement of 
instrumentation (and here no limit can be put on the development 
of science and technology) and on nature---even the most powerful 
and compact emission in the universe can be completely hidden by 
a thick enough screen.

\section{Summary}
\label{risalitisecsummary}

We have reviewed the main spectral properties of quasars, with 
an emphasis on continuum emission and a brief treatment of the 
main emission and absorption features. 

In the first part, we built an updated SED of quasars, mainly 
based on continuum emission of local ($z<0.4$) PG quasars.
The main improvements with respect to the previous reference 
work in the field, the atlas of Elvis et al.\ (1994), are: 
1) the treatment of selection effects in the X-rays, which 
allowed us to estimate the correct optical-to-X-ray flux
ratio for local quasars, 2) the extension of the SED in the UV 
at wavelengths shortward of the Lyman break, using {\em HST\/} 
observations of quasars, 3) the inclusion of the new IR data 
obtained with {\em ISO\/}, and 4) the use of a larger sample 
in the computation of the average spectral properties of quasars.

In the second part, we reviewed the main spectral properties of
obscured quasars. We also discussed the relation between dust and 
gas absorption, briefly describing the cases of X-ray obscured 
objects with no or little optical/near-IR absorption.

In the third and final part, we summarized the methods for 
finding obscured AGN in the universe, focusing on the problem 
of disentangling the active nucleus component in infrared luminous 
sources and disucssing the still open problems. Despite the improvements 
of the last few years, what we know is most likely only a fraction 
of the obscured quasars present in the universe. 

\begin{acknowledgments}
We wish to thank Amy Barger for giving us the opportunity to
contribute to this book, and for her really infinite patience
with our very long delays in preparing the manuscript.

\end{acknowledgments}


\begin{chapthebibliography}{1}

\bibitem[Antonucci \& Miller(1985)]{1985ApJ...297..621A} 
Antonucci, R. R. J., \& Miller, J. S.\ 1985, ApJ, 297, 621

\bibitem[Avni \& Tananbaum(1982)]{1982ApJ...262L..17A} 
Avni, Y., \& Tananbaum, H.\ 1982, ApJL, 262, L17

\bibitem[Avni \& Tananbaum(1986)]{1986ApJ...305...83A} 
Avni, Y., \& Tananbaum, H.\ 1986, ApJ, 305, 83 

\bibitem[Bechtold et al.(2003)]{2003ApJ...588..119B} 
Bechtold, J., et al.\ 2003, ApJ, 588, 119

\bibitem[Baldwin(1977)]{1977ApJ...214..679B} 
Baldwin, J. A.\ 1977, ApJ, 214, 679  

\bibitem[Barger et al.(1998)]{1998Nat...394..248B}
Barger, A. J., Cowie, L. L., Sanders, D. B., Fulton, E.,
Taniguchi, Y., Sato, Y., Kawara, K. \& Okuda, H.\ 1998, 
Nature, 394, 248

\bibitem[Barger et al.(2002)]{2002AJ....124.1839B} 
Barger, A. J., Cowie, L. L., Brandt, W. N., Capak, P., Garmire, G. P., 
Hornschemeier, A. E., Steffen, A. T., \& Wehner, E. H.\ 2002, AJ, 124, 1839 

\bibitem[Barger et al.(2003)]{2003AJ....126..632B} 
Barger, A. J., et al.\ 2003, AJ, 126, 632 

\bibitem[Braito]{Braito, 2004MNRAS, in press} 
Braito, V., et al.\ 2004, MNRAS, in press

\bibitem[Brandt, Laor, \& Wills(2000)]{2000ApJ...528..637B} 
Brandt, W. N., Laor, A., \& Wills, B. J.\ 2000, ApJ, 528, 637 

\bibitem[Brotherton et al.(2001)]{2001ApJ...546..775B} 
Brotherton, M. S., Tran, H. D., Becker, R. H., Gregg, M. D., 
Laurent-Muehleisen, S. A., \& White, R. L. 2001, ApJ, 546, 775

\bibitem[Brusa et al.(2003)]{2003A&A...409...65B} Brusa, M., et al.\ 2003, 
A\&A, 409, 65 

\bibitem[Chini, Kreysa, \& Biermann(1989)]{1989A&A...219...87C} 
Chini, R., Kreysa, E., \& Biermann, P. L.\ 1989, A\&A, 219, 87  

\bibitem[Clements et al.(2002)]{2002ApJ...581..974C} 
Clements, D. L., McDowell, J. C., Shaked, S., Baker, A. C., 
Borne, K., Colina, L., Lamb, S. A., \& Mundell, C.\ 2002, ApJ, 581, 974 

\bibitem[Crenshaw et al.(1999)]{1999ApJ...516..750C} 
Crenshaw, D. M., Kraemer, S. B., Boggess, A., Maran, S. P., 
Mushotzky, R. F., \& Wu, C.\ 1999, ApJ, 516, 750 

\bibitem[Croom et al.(2002)]{2002MNRAS.337..275C} 
Croom, S. M., et al.\ 2002, MNRAS, 337, 275 

\bibitem{a50} 
Cutri, R., et al.\ 2001, in ``The New Era of Wide Field
Astronomy'', Eds. R. Clowes, A. Adamson, \& G. Bromage 
(San Francisco: ASP Conference Series), 232, p78

\bibitem[Czerny \& Elvis(1987)]{1987ApJ...321..305C} 
Czerny, B., \& Elvis, M.\ 1987, ApJ, 321, 305 

\bibitem[de Grijp, Lub, \& Miley(1987)]{1987A&AS...70...95D} 
de Grijp, M. H. K., Lub, J., \& Miley, G. K.\ 1987, A\&AS, 70, 95 

\bibitem[Elitzur et al. 2004]{Elitzur, 2004, in press} 
Elitzur, M., Nenkova, M., \& Ivezic, Z., 2004, in
``The Neutral ISM in Starburst Galaxies'', in press (astro-ph/0309040) 

\bibitem[Elvis(1985)]{1985gecx.conf..291E} 
Elvis, M.\ 1985, in ``Galactic and Extra-Galactic Compact X-ray 
Sources'', Eds. Y. Tanaka \& W. H. G. Lewin, p291 

\bibitem[Elvis et al.(1978)]{1978MNRAS.183..129E}
Elvis, M., Maccacaro, T., Wilson, A. S., Ward, M. J.,
Penston, M. V., Fosbury, R. A. E., \& Perola, G. C.\ 1978,
MNRAS, 183, 129

\bibitem[Elvis, Risaliti, \& Zamorani(2002)]{2002ApJ...565L..75E} 
Elvis, M., Risaliti, G., \& Zamorani, G.\ 2002, ApJ, 565, L75

\bibitem[Elvis et al.(1994)]{1994ApJS...95....1E}
Elvis, M., et al. 1994, ApJS, 95, 1 (E94)

\bibitem[Fabbiano(1989)]{1989ARA&A..27...87F} 
Fabbiano, G.\ 1989, ARA\&A, 27, 87 

\bibitem[Fabian \& Iwasawa(1999)]{1999MNRAS.303L..34F}
Fabian, A. C., \& Iwasawa, K.\ 1999, MNRAS, 303, L

\bibitem[Fabian, Rees, Stella, \& White(1989)]{1989MNRAS.238..729F} 
Fabian, A. C., Rees, M. J., Stella, L., \& White, N. E.\ 1989, 
MNRAS, 238, 729 

\bibitem[Fabian et al.(2002)]{2002MNRAS.335L...1F} 
Fabian, A. C., et al.\ 2002, MNRAS, 335, L1 

\bibitem[Franceschini et al.(2003)]{2003MNRAS.343.1181F} 
Franceschini, A., et al.\ 2003, MNRAS, 343, 1181

\bibitem[Francis et al.(1991)]{1991ApJ...373..465F} 
Francis, P. J., Hewett, P. C., Foltz, C. B., Chaffee, F. H., 
Weymann, R. J., \& Morris, S. L.\ 1991, ApJ, 373, 465 

\bibitem[Genzel et al.(1998)]{1998ApJ...498..579G} 
Genzel, R., et al.\ 1998, ApJ, 498, 579 

\bibitem[George et al.(1998)]{1998ApJS..114...73G} 
George, I. M., Turner, T. J., Netzer, H., Nandra, K., 
Mushotzky, R. F., \& Yaqoob, T.\ 1998, ApJS, 114, 73 

\bibitem[George et al.(2000)]{2000ApJ...531...52G} 
George, I. M., Turner, T. J., Yaqoob, T., Netzer, H., Laor, A., 
Mushotzky, R. F., Nandra, K., \& Takahashi, T.\ 2000, ApJ, 531, 52 

\bibitem[Ghisellini, Haardt, \& Matt(1994)]{1994MNRAS.267..743G} 
Ghisellini, G., Haardt, F., \& Matt, G.\ 1994, MNRAS, 267, 743

\bibitem[Gilli, Salvati, \& Hasinger(2001)]{2001A&A...366..407G} 
Gilli, R., Salvati, M., \& Hasinger, G.\ 2001, A\&A, 366, 407

\bibitem[Guainazzi et al.(2000)]{2000A&A...356..463G} 
Guainazzi, M., Matt, G., Brandt, W. N., Antonelli, L. A., 
Barr, P., \& Bassani, L.\ 2000, A\&A, 356, 463

\bibitem[Haas et al.(2003)]{2003A&A...402..87H} 
Haas, M., et al. 2003, A\&A, 402, 87 

\bibitem[Hagen, Groote, Engels, \& Reimers(1995)]{1995A&AS..111..195H} 
Hagen, H.-J., Groote, D., Engels, D., \& Reimers, D.\ 1995, A\&AS, 111, 195 

\bibitem[Hughes et al.(1998)]{1998Natur...394..241H}
Hughes, D. H., et al.\ 1998, Nature, 394, 241

\bibitem[Imanishi \& Dudley(2000)]{2000ApJ...545..701I} 
Imanishi, M., \& Dudley, C. C.\ 2000, ApJ, 545, 701 

\bibitem[Kellermann et al.(1989)]{1989AJ.....98.1195K} 
Kellermann, K. I., Sramek, R., Schmidt, M., Shaffer, D. B., \& 
Green, R.\ 1989, AJ, 98, 1195

\bibitem[Kim \& Sanders(1998)]{1998ApJS..119...41K}
Kim, D.-C., \& Sanders, D. B.\ 1998, ApJS, 119, 41

\bibitem[Krongold et al.(2003)]{2003ApJ...597..832K} 
Krongold, Y., Nicastro, F., Brickhouse, N. S., Elvis, M., 
Liedahl, D. A., \& Mathur, S.\ 2003, ApJ, 597, 832

\bibitem[Kuraszkiewicz et al.(2003)]{2003ApJ...590..128K} Kuraszkiewicz, 
J.~K., et al.\ 2003, ApJ, 590, 128 

\bibitem[Laor(1991)]{1991ApJ...376...90L} 
Laor, A.\ 1991, ApJ, 376, 90  

\bibitem[Laor et al.(1997)]{1997ApJ...477...93L} 
Laor, A., Fiore, F., Elvis, M., Wilkes, B. J., \& McDowell, J. C.\ 1997, 
ApJ, 477, 93 

\bibitem[Laurent et al.(2000)]{2000A&A...359..887L} 
Laurent, O., Mirabel, I. F., Charmandaris, V., Gallais, P., 
Madden, S. C., Sauvage, M., Vigroux, L., \& Cesarsky, C.\ 2000, A\&A, 
359, 887 

\bibitem[Lonsdale, Smith, \& Lonsdale(1993)]{1993ApJ...405L...9L} 
Lonsdale, C. J., Smith, H. J., \& Lonsdale, C. J.\ 1993, ApJ, 405, L9 

\bibitem[Lutz et al.(2003)]{2003A&A...409..867L} 
Lutz, D., Sturm, E., Genzel, R., Spoon, H. W. W., Moorwood, A. F. M., 
Netzer, H., \& Sternberg, A.\ 2003, A\&A, 409, 867

\bibitem[Maccacaro, Perola, \& Elvis(1982)]{1982ApJ...257...47M} 
Maccacaro, T., Perola, G. C., \& Elvis, M.\ 1982, ApJ, 257, 47

\bibitem[Maiolino et al.(2001)]{2001A&A...365...28M}
Maiolino, R., Marconi, A., Salvati, M., Risaliti, G.,
Severgnini, P., Oliva, E., La Franca, F., \& Vanzi, L.\ 2001, A\&A,
365, 28

\bibitem[Maiolino \& Rieke(1995)]{1995ApJ...454...95M} 
Maiolino, R., \& Rieke, G. H.\ 1995, ApJ, 454, 95

\bibitem[Malkan \& Sargent(1982)]{1982ApJ...254...22M} 
Malkan, M. A., \& Sargent, W. L. W.\ 1982, ApJ, 254, 22  

\bibitem[Matt et al.(1997)]{1997A&A...325L..13M} 
Matt, G., et al.\ 1997, A\&A, 325, L13 

\bibitem[Mineo et al.(2000)]{2000A&A...359..471M} 
Mineo, T., et al.\ 2000, A\&A, 359, 471 

\bibitem[Nagar, Falcke, Wilson, \& Ho(2000)]{2000ApJ...542..186N} 
Nagar, N. M., Falcke, H., Wilson, A. S., \& Ho, L. C.\ 2000, ApJ, 542, 186

\bibitem[Nagar et al.(2003)]{2003A&A...409..115N} 
Nagar, N. M., Wilson, A. S., Falcke, H., Veilleux, S., \& 
Maiolino, R.\ 2003, A\&A, 409, 115 

\bibitem[Nandra et al.(1997)]{1997ApJ...477..602N} 
Nandra, K., George, I. M., Mushotzky, R. F., Turner, T. J., \& 
Yaqoob, T.\ 1997, ApJ, 477, 602

\bibitem[Neufeld, Maloney, \& Conger(1994)]{1994ApJ...436L.127N} 
Neufeld, D. A., Maloney, P. R., \& Conger, S.\ 1994, ApJ, 436, L127 

\bibitem[Panessa \& Bassani(2002)]{2002A&A...394..435P} 
Panessa, F., \& Bassani, L.\ 2002, A\&A, 394, 435

\bibitem[Pendleton et al.(1994)]{1994ApJ...437..683P} 
Pendleton, Y. J., Sandford, S. A., Allamandola, L. J., 
Tielens, A. G. G. M., \& Sellgren, K.\ 1994, ApJ, 437, 683

\bibitem[Perola et al.(2002)]{2002A&A...389..802P} 
Perola, G. C., Matt, G., Cappi, M., Fiore, F., Guainazzi, M., 
Maraschi, L., Petrucci, P. O., \& Piro, L.\ 2002, A\&A, 389, 802  

\bibitem[Risaliti(2002)]{2002A&A...386..379R} 
Risaliti, G.\ 2002, A\&A, 386, 379

\bibitem[Risaliti, Elvis, Gilli, \& Salvati(2003)]{2003ApJ...587L...9R} 
Risaliti, G., Elvis, M., Gilli, R., \& Salvati, M.\ 2003a, ApJ, 587, L9 

\bibitem[Risaliti, Gilli, Maiolino, \& Salvati(2000)]{2000A&A...357...13R}
Risaliti, G., Gilli, R., Maiolino, R., \& Salvati, M.\ 2000, A\&A, 357, 13

\bibitem[Risaliti et al.(2001)]{2001A&A...371...37R}
Risaliti, G., Marconi, A., Maiolino, R., Salvati, M., \&
Severgnini, P.\ 2001, A\&A, 371, 37

\bibitem[Risaliti, Woltjer, \& Salvati(2003)]{2003A&A...401..895R} 
Risaliti, G., Woltjer, L., \& Salvati, M.\ 2003b, A\&A, 401, 895 

\bibitem[Risaliti et al.(2003)]{2003ApJ...595L..17R}
Risaliti, G., et al.\ 2003c, ApJ, 595, L17

\bibitem[Sanders \& Mirabel(1996)]{1996ARA&A...34..749S}
Sanders, D. B., \& Mirabel, I. F.\ 1996, ARA\&A, 34, 749

\bibitem[Sanders et al.(1989)]{1989ApJ...347...29S} 
Sanders, D. B., Phinney, E. S., Neugebauer, G., Soifer, B. T., \& 
Matthews, K.\ 1989, ApJ, 347, 29 

\bibitem[Schmidt \& Green(1983)]{1983ApJ...269..352S} 
Schmidt, M., \& Green, R. F.\ 1983, ApJ, 269, 352 

\bibitem[Schneider, Schmidt, \& Gunn(1994)]{1994AJ....107.1245S} 
Schneider, D. P., Schmidt, M., \& Gunn, J. E.\ 1994, AJ, 107, 1245

\bibitem[Shields(1978)]{1978Natur.272..706S} 
Shields, G. A.\ 1978, Nature, 272, 706 

\bibitem[Smail et al.(1997)]{1997ApJ...490L..S}
Smail, I., Ivison, R. J., \& Blain, A. W.\ 1997, 490, L5

\bibitem[Smith, Lonsdale, \& Lonsdale(1998)]{1998ApJ...492..137S}
Smith, H. E., Lonsdale, C. J., \& Lonsdale, C. J.\ 1998, ApJ, 492, 137

\bibitem[Tanaka et al.(1995)]{1995Natur.375..659T} 
Tanaka, Y., et al.\ 1995, Nature, 375, 659 

\bibitem[Telfer, Zheng, Kriss, \& Davidsen(2002)]{2002ApJ...565..773T} 
Telfer, R. C., Zheng, W., Kriss, G. A., \& Davidsen, A. F.\ 2002, ApJ, 
565, 773 

\bibitem[Tokunaga et al.(1991)]{1991ApJ...380..452T} 
Tokunaga, A. T., Sellgren, K., Smith, R. G., Nagata, T., 
Sakata, A., \& Nakada, Y.\ 1991, ApJ, 380, 452 

\bibitem[Vanden Berk et al.(2001)]{2001AJ....122..549V} 
Vanden Berk, D. E., et al.\ 2001, AJ, 122, 549 

\bibitem[Vaughan \& Fabian 2004]{Vaughan, 2004MNRAS, in press} 
Vaughan, S., \& Fabian, A.C.\ 2004, MNRAS, in press (astro-ph/0311473)

\bibitem[Veilleux, Kim, \& Sanders(1999)]{1999ApJ...522..113V}
Veilleux, S., Kim, D.-C., \& Sanders, D. B.\ 1999, ApJ, 522, 113

\bibitem[Veilleux \& Osterbrock(1987)]{1987ApJS...63..295V} 
Veilleux, S., \& Osterbrock, D. E.\ 1987, ApJS, 63, 295 

\bibitem[V{\' e}ron-Cetty, Woltjer, Staveley-Smith, 
\& Ekers(2000)]{2000A&A...362..426V} 
V{\' e}ron-Cetty, M.-P., Woltjer, L., 
Staveley-Smith, L., \& Ekers, R. D.\ 2000, A\&A, 362, 426

\bibitem[Vignali, Brandt, \& Schneider(2003)]{2003AJ....125..433V} 
Vignali, C., Brandt, W. N., \& Schneider, D. P.\ 2003, AJ, 125, 433 

\bibitem[Vignati et al.(1999)]{1999A&A...349L..57V} 
Vignati, P., et al.\ 1999, A\&A, 349, L57

\bibitem[White et al.(2000)]{2000ApJS..126..133W} 
White, R. L., et al.\ 2000, ApJS, 126, 133  

\bibitem[Wilkes et al.(2002)]{2002ApJ...564L..65W}
Wilkes, B. J., Schmidt, G. D., Cutri, R. M., Ghosh, H.,
Hines, D. C., Nelson, B., \& Smith, P. S.\ 2002, ApJ, 564, L65

\bibitem[Wilkes et al.(1994)]{1994ApJS...92...53W} 
Wilkes, B. J., Tananbaum, H., Worrall, D. M., Avni, Y., 
Oey, M. S., \& Flanagan, J.\ 1994, ApJS, 92, 53  

\bibitem[Wills, Netzer, \& Wills(1985)]{1985ApJ...288...94W} 
Wills, B. J., Netzer, H., \& Wills, D.\ 1985, ApJ, 288, 94 

\bibitem[Wisotzki, Koehler, Groote, \& Reimers(1996)]{1996A&AS..115..227W} 
Wisotzki, L., Koehler, T., Groote, D., \& Reimers, D.\ 1996, A\&AS, 115, 
227 

\bibitem[Yuan, Brinkmann, Siebert, \& Voges(1998)]{1998A&A...330..108Y} 
Yuan, W., Brinkmann, W., Siebert, J., \& Voges, W.\ 1998, A\&A, 330, 108 

\bibitem[Yuan, Siebert, \& Brinkmann(1998)]{1998A&A...334..498Y} 
Yuan, W., Siebert, J., \& Brinkmann, W.\ 1998, A\&A, 334, 498 

\bibitem[Zamorani et al.(1981)]{1981ApJ...245..357Z} 
Zamorani, G., et al.\ 1981, ApJ, 245, 357 

\end{chapthebibliography}
\end{document}